\documentclass[aps,pre,reprint,superscriptaddress]{revtex4-2}
\usepackage{graphicx}
\usepackage{amsmath}
\usepackage{latexsym}
\usepackage{amsthm, amscd, amsfonts, amssymb}

\begin{document}

\title{Controlled nonautonomous matter-wave solitons in spinor Bose-Einstein
condensates with spatiotemporal modulation }
\author{Cui-Cui Ding}
\author{Qin Zhou}
\email{qinzhou@whu.edu.cn}
\affiliation{Research Center of Nonlinear Science, School of Mathematical and Physical Sciences, \\Wuhan Textile University, Wuhan 430200, China}
\author{Si-Liu Xu}
\affiliation{School of Biomedical Engineering and Medical Imaging, Xianning Medical College, Hubei University of Science and Technology, Xianning 437100, China}
\author{Yun-Zhou Sun}
\email{syz@wtu.edu.cn}
\affiliation{Research Center of Nonlinear Science, School of Mathematical and Physical Sciences, \\Wuhan Textile University, Wuhan 430200, China}
\author{Wen-Jun Liu}
\email{jungliu@bupt.edu.cn}
\affiliation{State Key Laboratory of Information Photonics and Optical Communications, School of Science, Beijing University of Posts and Telecommunications, P. O. Box 122, Beijing 100876, China}
\author{Dumitru Mihalache}
\affiliation{Horia Hulubei National Institute of Physics and Nuclear Engineering, 077125 Magurele, Bucharest, Romania}
\author{Boris A. Malomed}
\affiliation{Department of Physical Electronics, School of Electrical Engineering, Faculty of Engineering, Tel Aviv University, Tel Aviv 69978, Israel}
\affiliation{Instituto de Alta Investigación, Universidad de Tarapacá, Casilla 7D, Arica, Chile}
\date{\today }

\begin{abstract}
To study controlled evolution of nonautonomous matter-wave solitons in
spinor Bose-Einstein condensates with spatiotemporal modulation, we focus on
a system of three coupled Gross-Pitaevskii (GP) equations with
space-time-dependent external potentials and temporally modulated gain/loss
distributions. An integrability condition and a nonisospectral Lax pair for
the coupled GP equations are obtained. Using it, we derive an infinite set
of dynamical invariants, the first two of which are the mass and momentum.
The Darboux transform is used to generate one- and two-soliton solutions.
Under the action of different external potentials and gain/loss
distributions, various solutions for controlled nonautonomous matter-wave
solitons of both ferromagnetic and polar types are obtained, such as
self-compressed, snake-like and stepwise solitons, and as well as breathers.
In particular, the formation of states resembling rogue waves,
under the action of a sign-reversible gain-loss distribution, is
demonstrated too. Shape-preserving and changing interactions between two
nonautonomous matter-wave solitons and bound states of solitons are
addressed too. In this context, spin switching arises in the
polar-ferromagnetic interaction. Stability of the nonautonomous matter-wave
solitons is verified by means of systematic simulations of their perturbed
evolution.
\end{abstract}

\maketitle










\section{Introduction}


For decades, Bose-Einstein condensates (BECs) of ultracold atoms have been
studied extensively since their creation in the first experiments \cite%
{Anderson1995,Hulet,Ketterle,Pitaevskii2016,Malomed2022,Kengne2021,ZhangChen2021,Musolino2022,Henderson2022}%
. The great interest in this topic has been driven, in particular, by two
beneficial features: (1) intrinsic properties of the system, such as the
strength of the interatomic interactions, can be manipulated by dint of
magnetic fields and lasers; (2) the mean-field theory provides very accurate
description of BEC in dilute atomic gases \cite%
{Pitaevskii2016,Pethick,Kengne2021,ZhangChen2021,Musolino2022,Henderson2022}.

In particular, broad attention has been attracted to spinor
(multi-component) BECs maintained by optical traps \cite%
{Kawaguchi2012,Kartashov2019,Malomed2019a,Mihalache2021,LiLi2005,LiZhu2020,Stamper-Kurn2013,Bersano2018,Evrard2021,Kim K2021}%
, as their internal spin degrees of freedom give rise to abundant phenomena,
including magnetic crystallization, spin textures and fractional vortices,
which have no counterparts in the magnetically trapped condensates with the
spin degree being frozen~\cite{Evrard2021,Kim K2021,Vengalattore2008}. Many
experimental and theoretical studies of spinor BECs have revealed a variety
of interesting phenomena, such as polar-to-ferromagnetic phase transitions,
quantum knots, condensation of magnon excitations, and various kinds of
nonlinear excitations consisting of dark/bright solitons, soliton complexes,
rogue waves, vortices, etc.~\cite{Borgh M O2017,Ollikainen2017}. Matter-wave
solitons in atom optics may be used in the design of atom lasers, atom
interferometry and coherent atom transport~\cite{Meystre2001,Sekh G
A2015,Peter}. Matter-wave solitons with internal spin degrees of freedom may
find still more diverse application~\cite{Chai2020a,Chai2021a}. In the
mean-field approximation, the spinor BEC can be described by a set of
multicomponent Gross-Pitaevskii (GP) equations \cite{Kawaguchi2012}.

Recently, there has been increased interest in studying spatiotemporally
modulated BEC system with time-space-dependent external potentials~\cite%
{Zhang2009,Rajendrana2010,Yao Y Q2018}, time-variable gain-loss distribution
provided by optical pumping or depletion~\cite{Atre2006}, and time-dependent
nonlinearity manipulated by dint of the Feshbach-resonance technique~\cite%
{Frantz,Enomoto2008,Yan2013,Shen2014}. In particular, the time-dependent
terms in the corresponding GP equations can generate various results in the
framework of the dynamical management of solitons~\cite{Malomed2006}. The
celebrated instances are the dispersion management in fiber optics~\cite%
{Malomed2006,Turitsyn2012} and nonlinearity management in BEC through the
Feshbach resonance technique~\cite{Frantz,Yan2013,Enomoto2008}.

In spatiotemporally modulated systems, nonautonomous solitons, which
propagate with varying amplitudes and velocities, can be obtained in various
physical settings, including hydrodynamics, nonlinear optics, matter waves,
etc.~\cite{Shen2014,Serkin2007,Kengne2021}. In particular, nonautonomous
matter waves in spatiotemporally modulated spinor BECs feature properties
different from those of classical matter waves \cite%
{Serkin2007,Rajendran2011,YangZhao2011,Wang D S2013}.

This paper addresses two aspects. First, we consider integrability
conditions for the nonautonomous spin-1 BEC system with a spatiotemporally
modulated external potential and time-varying gain-loss distribution, and
then derive the respective Lax pair and an infinite set of conservation
laws. Second, we construct an $N$-th-order Darboux transformation and study
the dynamics of nonautonomous matter-wave solitons of both the ferromagnetic
and polar states for the nonautonomous spin-1 BEC system with several
different kinds of space-time-dependent external potentials and time-varying
gain-loss patterns. Bound states of solitons, and shape-preserving and
shape-changing interactions between them are addressed too.

The paper is organized as follows. In Sec. II, the model is formulated for
the spinor BEC with the spatiotemporal modulation. In Sec. III, we first
derive the integrability condition and nonisospectral Lax pair for the
coupled GP equations. Then we derive the infinitely set of conservation laws
and put forward the physical meaning of the two lowest-order ones. The $N$%
-th-order Darboux transform is also derived. In Sec. IV, various controlled
nonautonomous matter-wave solitons of both ferromagnetic and polar types are
obtained for different external potentials and gain-loss profiles. We also
analyze stability of the nonautonomous matter-wave solitons by means of
numerical simulations. In Sec. V, shape-preserving and shape-changing
interactions between two nonautonomous matter-wave solitons and bound-states
of the solitons are addressed. Conclusions are formulated in Sec. VI.

\section{The model}

In the present work, we focus on the dynamics of the spinor BEC with a
spatiotemporally-dependent external harmonic-oscillator (HO) potential and
time-variable atom gain-loss distributions. Here, we consider the quasi-one
dimensional regime: the cigar-shaped trap is elongated in the $x$ direction
and strongly confined in the transverse directions $y$ and $z$, which is
available to the experiment~\cite{Carr2004,Davidson}. In the $F=1$ state,
the distribution of atoms is presented by the three-component macroscopic
BEC wave function: $\mathbf{\Phi }(x,t)\equiv \lbrack \Phi _{+1}(x,t),\Phi
_{0}(x,t),\Phi _{-1}(x,t)]$, with the three components pertaining to the
three internal states $m_{F}=+1,0,-1$, where $m_{F}$ is the magnetic quantum
number. The dynamics of the spinor BEC under the action of
spatiotemporally-dependent HO potentials and time-variable gain-loss
distributions is governed by the following coupled GP equations within the
mean field approximation~\cite%
{Kawaguchi2012,Ieda2004,Stamper-Kurn2013,Bersano2018}:
\begin{widetext}
\begin{subequations} \label{system0}
\begin{align}
\text{i}\hbar\Phi_{\pm1,t}&=-\frac{\hbar^2}{2M}\Phi_{\pm1,xx}
+(c_0+c_2)(|\Phi_{\pm1}|^2+|\Phi_{0}|^2)\Phi_{\pm1}
+(c_0-c_2)|\Phi_{\mp1}|^2\Phi_{\pm1}+c_2\Phi_0^2\Phi_{\mp1}^*
+V_{\text{ext}}(x,t)\Phi_{\pm1},\\
\text{i}\hbar\Phi_{0,t}&=-\frac{\hbar^2}{2M}\Phi_{0,xx}
+(c_0+c_2)(|\Phi_{+1}|^2+|\Phi_{-1}|^2)\Phi_{0}
+c_0|\Phi_{0}|^2\Phi_0+2c_2\Phi_{0}^*\Phi_{+1}\Phi_{-1}
+V_{\text{ext}}(x,t)\Phi_{0},
\end{align}
\end{subequations}
\end{widetext}where $V_{\text{ext}}(x,t)$ represents the spatiotemporally
modulated external potential and time-dependent gain-loss distribution, $%
\ast $ stands for the complex conjugate, and $M$ is the atomic mass.
Further, $c_{0}=\left( g_{0}+2g_{2}\right) /3$ and $c_{2}=\left(
g_{2}-g_{0}\right) /3$ stand, respectively, for the effective constants of
the spin-preserving and spin-exchange interaction,
\begin{equation}
g_{f}=\frac{4\hbar ^{2}a_{f}}{Ma_{\bot }^{2}(1-Ca_{f}/a_{\bot })}
\end{equation}%
$(f=0,2)$ denote effective coupling constants, and $a_{f}$ is the $s$-wave
scattering length in the channel with the total hyperfine spin $f$. Next, $%
a_{\bot }$ is the transverse size of the ground state. In the present work,
we address the case of $c_{0}=c_{2}=-c<0$, hence $2g_{0}=-g_{2}>0$, which
represents the attractive spin-preserving and ferromagnetic spin-exchange
interactions. Then, through rescaling $\mathbf{\Phi }\rightarrow (\phi _{+1},%
\sqrt{2}\phi _{0},\phi _{-1})$ casts system~(\ref{system0}) in the form of

\begin{subequations}
\label{system}
\begin{eqnarray}
\text{i}\phi _{\pm 1,t} &=&-\phi _{\pm 1,xx}-2(|\phi _{\pm 1}|^{2}+2|\phi
_{0}|^{2})\phi _{\pm 1}  \notag \\
&&-2\phi _{0}^{2}\phi _{\mp 1}^{\ast }-v_{\text{ext}}(x,t)\phi _{\pm 1},
\end{eqnarray}%
\begin{eqnarray}
\text{i}\phi _{0,t} &=&-\phi _{0,xx}-2(|\phi _{+1}|^{2}+|\phi
_{0}|^{2}+|\phi _{-1}|^{2})\phi _{0}  \notag \\
&&-2\phi _{-1}\phi _{0}^{\ast }\phi _{+1}-v_{\text{ext}}(x,t)\phi _{0},
\end{eqnarray}%
where the coordinates and time are, measured, respectively, in units of $%
\hbar \sqrt{a_{\bot }/\left( 2Mc\right) }$ and $\hbar a_{\bot }/c$,, and
\end{subequations}
\begin{equation}
v_{\text{ext}}(x,t)\equiv U_{\text{trap}}(x,t)+\text{i}\Gamma (t).
\label{vext}
\end{equation}%
Here
\begin{equation}
U_{\text{trap}}(x,t)=U_{p}(t)x^{2}+\gamma (t)x  \label{UUU}
\end{equation}%
is the temporally modulated trapping potential~\cite{Serkin2007,Serkin2010},
and $\Gamma (t)$ stands for the time-dependent coefficient of the atomic gain and loss, which can be implemented, severally, by loading atoms into the BEC with the optical pump and an electron beam or a strongly focused resonant blast laser in the BEC~\cite{Janis2005,Gericke2008,Wurtz2009}.

\section{The Lax pair, Darboux transform, and the infinite set of
conservation laws}

In this section, we aim to derive an integrability condition for system (\ref%
{system}) and construct the respective Lax pair and infinite set of
dynamical invariants (conservation laws). As system~(\ref{system}) is
nonautonomous with the spatiotemporal modulation, to derive the Lax pair, we
utilize the generalized Ablowitz-Kaup-Newell-Segur formalism~\cite%
{Ablowitz1973} and attempt to construct a nonisospectral Lax pair for
system~(\ref{system}) as

\begin{equation}
\Psi _{x}=\mathbf{U}\Psi ,~~\Psi _{t}=\mathbf{V}\Psi  \label{lax pair}
\end{equation}%
where $\mathbf{U}=\text{i}\lambda (t)J+P$ and $\mathbf{V}=2\text{i}\lambda
(t)^{2}J+2\lambda (t)V_{1}+\text{i}V_{0}$, $\lambda (t)$ is a complex
nonisospectral parameter,
\begin{equation}
\Psi =(\mathcal{H},\mathcal{Y})^{T}  \label{Jost}
\end{equation}%
is the matrix Jost function, $\mathcal{H}$ and $\mathcal{Y}$ are $2\times 2$
matrices, and other matrices are expressed as
\begin{equation}
\begin{aligned} J&=\begin{pmatrix} -I & O \\ O & I \end{pmatrix},~~
P=\begin{pmatrix} O & Q \\ -Q^{\dag} & O \end{pmatrix},\\
V_0&=\begin{pmatrix} QQ^{\dag} & Q_x+2\text{i}\Gamma(t)xQ \\
Q^{\dag}_x-2\text{i}\Gamma(t)xQ^{\dag} & -Q^{\dag}Q \end{pmatrix},\\
V_1&=-\text{i}\Lambda(x,t)J+
P,~\Lambda(x,t)=\left[\Gamma(t)+\frac{\gamma(t)}{4\lambda(t)}\right]x,\\
\lambda(t)&=\xi e^{-2\int \Gamma(t)\, dt}-\frac{1}{2}e^{-2\int \Gamma(t)\,
dt} \int \gamma(t)e^{2\int \Gamma(t)\, dt}\, dt,\\
Q&=e^{-\frac{\text{i}\Gamma(t)x^2}{2}} \begin{pmatrix} \phi_{+1} & \phi_0 \\
\phi_0 & \phi_{-1} \end{pmatrix}. \end{aligned}  \label{qmatrix}
\end{equation}%
Here $I$ and $O$ are the $2\times 2$ unity and zero matrices,
\textquotedblleft $\dag $\textquotedblright\ stands for the Hermitian
conjugate, and $\xi $ is a complex constant. The Lax pair~(\ref{lax pair})
is valid under the following integrability condition imposed on the
time-dependent coefficients in potentials (\ref{vext}) and (\ref{UUU}):
\begin{equation}
U_{p}(t)=(1/2)\Gamma _{t}(t)+\Gamma ^{2}(t),  \label{condition}
\end{equation}%
while $\gamma (t)$ remains an arbitrary real function of $t$. Here, we
directly find the integrability condition (\ref{condition}) for the
nonautonomous system~(\ref{system}), and then solve the integrable system~(%
\ref{system}) analytically, by dint of the Darboux transformation.
Alternatively, it may be possible to transform the integrable version of the
nonautonomous system~into the autonomous integrable one discovered in Ref.
\cite{Ieda2004} through an appropriate transformation of $\Phi $ and $x,t$
(as it could be done with many other models \cite{Kengne2021}), although the
latter approach appears to be quite cumbersome for the present system,
therefore it is not pursued here.

As mentioned above, in the case of the time-modulated system spectral
parameter $\lambda (t)$ is not a constant but a function of $t$, which is
determined by coefficients $\Gamma (t)$ and $\gamma (t)$ that, respectively,
account for the gain-loss term and linear potential in the system, and
complex constant $\xi $. The time-dependent spectral parameter has a
significant impact on the nonautonomous matter-wave solitons, which is
considered in detail in the following sections.

An important consequence of the integrability of system~(\ref{system}) with
Lax pair~(\ref{lax pair}) is that it possesses an infinite set of
conservation laws. To derive them, we define an auxiliary matrix function $%
\Upsilon =\mathcal{H}\mathcal{Y}^{-1}$, in terms of components of the Jost
function (\ref{Jost}). By substituting $\Upsilon $ it into the Lax pair~(\ref%
{lax pair}), the following Riccati-type equation can be obtained:
\begin{equation}
\Upsilon _{x}=Q-2\text{i}\lambda (t)\Upsilon +\Upsilon Q^{\dag }\Upsilon ,
\label{Riccati-type eq}
\end{equation}%
where $Q$ is given by Eq.~(\ref{qmatrix}). Taking the expansion
\begin{equation}
Q^{\dag }\Upsilon =\sum_{k=1}^{\infty }\frac{\Upsilon _{k}}{\lambda (t)^{k}},
\label{expansion}
\end{equation}%
where $\Upsilon _{k}$ ($k=1,2,3,\ldots $) are $2\times 2$ matrix functions
of $x$ and $t$. Substituting Eq.~(\ref{expansion}) in Eq.~(\ref{Riccati-type
eq}) and equating the respective net coefficients in front of each power of $%
\lambda (t)$ to zero, one derives the following recurrence relations:
\begin{subequations}
\label{recurrence relations}
\begin{align}
\Upsilon _{1}=& -\frac{\text{i}}{2}Q^{\dag }Q, \\
\Upsilon _{2}=& -\frac{\text{i}}{2}\left[ Q_{x}^{\dag }(Q^{\dag
})^{-1}\Upsilon _{1}-\Upsilon _{1,x}\right] =\frac{1}{4}Q^{\dag }Q_{x}, \\
\Upsilon _{3}=& -\frac{\text{i}}{2}\left[ Q_{x}^{\dag }(Q^{\dag
})^{-1}\Upsilon _{2}-\Upsilon _{2,x}+\Upsilon _{1}\Upsilon _{1}\right] ,
\notag \\
=& \frac{\text{i}}{8}(Q^{\dag }Q_{xx}+Q^{\dag }QQ^{\dag }Q), \\
\vdots &  \notag \\
\Upsilon _{k+1}=& -\frac{\text{i}}{2}\left[ Q_{x}^{\dag }(Q^{\dag
})^{-1}\Upsilon _{k}-\Upsilon _{k,x}+\sum_{j=1}^{k-1}\Upsilon _{j}\Upsilon
_{k-j}\right] . \\
& (k=2,3,4,\ldots )  \notag
\end{align}%
Then, taking into account the compatibility condition $(\ln \mathcal{Y}%
)_{xt}=(\ln \mathcal{Y})_{tx}$, and utilizing Eqs.~(\ref{expansion}) and~(%
\ref{recurrence relations}), we derive the infinite set of conservation laws
for system~(\ref{system}) in the form of
\end{subequations}
\begin{equation}
\frac{\partial U_{j}}{\partial t}=\frac{\partial F_{j}}{\partial x}%
,~~(j=1,2,3,\ldots )  \label{conservation laws}
\end{equation}%
with
\begin{equation}
\begin{aligned} U_1=& ~2\text{i}\Upsilon_1=Q^{\dag}Q,\\ F_1=&
-2\text{i}\left[-2\Upsilon_2+2\Gamma(t)x\Upsilon_1+\text{i}Q^{\dag}_x(Q^{%
\dag})^{-1}\Upsilon_1\right],\\ U_2=& ~4\Upsilon_2=Q^{\dag}Q_x,\\ F_2=& ~
(-2\text{i})^2\left[-2\Upsilon_3+2\Gamma(t)x\Upsilon_2+\text{i}Q^{%
\dag}_x(Q^{\dag})^{-1}\Upsilon_2\right],\\ \vdots & \\ U_j=& -(-2\text{i})^j
\Upsilon_j,\\ U_j=&
(-2\text{i})^j\left[-2\Upsilon_{j+1}+2\Gamma(t)x\Upsilon_j+\text{i}Q^{%
\dag}_x(Q^{\dag})^{-1}\Upsilon_j\right], \end{aligned}
\label{conservation laws2}
\end{equation}%
where $U_{j}$ and $F_{j}$ denote the conserved densities and respective
fluxes, respectively.

The integrable form of system~(\ref{system}) without the external potential,
$v_{\text{ext}}(x,t)=0$, and with constant coefficients has been
investigated in Ref.~\cite{Ieda2004}, where dynamical invariants were
derived. Here, we identify the physical purport of the first few
conservation laws obtained here and put forward relations between
conservation laws~(\ref{conservation laws}) and those derived in Ref.~\cite%
{Ieda2004} for the integrable system with constant coefficients. According
to Eqs.~(\ref{conservation laws}), we have the following conserved
quantities for system~(\ref{system}):
\begin{subequations}
\label{conserved quantities}
\begin{align}
I_{1}=& \int dx\,\text{tr}(U_{1})=\int dx\,\text{tr}\{Q^{\dag }Q\}, \\
I_{2}=& \int dx\,\text{tr}(U_{2})=\int dx\,\text{tr}\{Q^{\dag }Q_{x}\},
\end{align}%
which are related to the conserved quantities including total number of
atoms in the condensate and its total momentum, derived in Ref.~\cite%
{Ieda2004}:
\end{subequations}
\begin{subequations}
\label{conserved quantities2}
\begin{align}
\mathbf{total~number:}~~N_{T}=& \int dx\,\mathbf{\Phi ^{\dag }}\cdot \mathbf{%
\Phi }  \notag \\
=& \int dx\,\text{tr}\{Q^{\dag }Q\}=I_{1}, \\
\mathbf{total~momentum:}~~P_{T}=& \int dx\,(-\text{i}\hbar \,\mathbf{\Phi
^{\dag }}\cdot \partial _{x}\mathbf{\Phi })  \notag \\
=& \int dx\,[-\text{i}\hbar \,\text{tr}\{Q^{\dag }Q_{x}\}]  \notag \\
=& -\text{i}\hbar \,I_{2}
\end{align}%
where \textquotedblleft $\text{tr}$\textquotedblright\ represents the matrix
trace. These relations offer the physical identification of the first two
dynamical invariants in Eq.~(\ref{conservation laws}) as the mass and
momentum, respectively. As well as in other integrable systems, the physical
interpretation of the higher conserved quantities is not straightforward.

The Darboux transform (DT) is an effective method to construct analytical
solutions for the nonlinear evolution equations \cite{Darboux}. We have
derived the DT for system~(\ref{system}) based on Lax pair~(\ref{lax pair})
and adopted the obtained DT to construct exact solutions of system (\ref%
{system}) for nonautonomous matter-wave solitons. Let $\Psi _{1}=(\mathcal{H}%
_{1}^{[0]},\mathcal{Y}_{1}^{[0]})^{T}$ be a zero-order complex matrix
solution of Lax pair~(\ref{lax pair}) with $Q=Q[0]$ and $\lambda (t)=\lambda
_{1}(t)$. Then the DT is used to construct the first-order solution as
\end{subequations}
\begin{subequations}
\label{DT1}
\begin{align}
\Psi \lbrack 1]=& T[1]\Psi ,~~T[1]=\tau _{1}(\lambda )[\lambda I-H[0]\Lambda
_{1}H[0]^{-1}], \\
Q[1]=& Q[0]-2\text{i}(\lambda _{1}-\lambda _{1}^{\ast })(\mathcal{Y}%
_{1}^{[0]}{\mathcal{H}_{1}^{[0]}}^{-1}+\mathcal{H}_{1}^{[0]\ast }{\mathcal{Y}%
_{1}^{[0]\ast }}^{-1})^{-1},
\end{align}%
with
\end{subequations}
\begin{equation}
\begin{aligned}
\tau_1(\lambda)=&[\text{det}(\lambda(t)I-H[0]\Lambda_1H[0]^{-1})]^{-%
\frac{1}{2}}\\
=&[(\lambda(t)-\lambda_1(t))(\lambda(t)-\lambda_1^*(t))]^{-\frac{1}{2}},\\
H[0]=&\begin{pmatrix} \mathcal{H}_1^{[0]} & -\mathcal{Y}_1^{[0]} \\
\mathcal{Y}_1^{[0]} & \mathcal{H}_1^{[0]} \end{pmatrix},~~
\Lambda_1=\begin{pmatrix} \lambda_1(t)I & O \\ O & \lambda_1^*(t)I
\end{pmatrix}. \end{aligned}
\end{equation}%
The above results take advantage of the Hermitian-symmetric profile of
matrix $Q$ (i.e., $Q^{\dag }=Q^{\ast }$). Making use of this profile, it can
be verified that if $\Psi _{1}=(\mathcal{H}_{1}^{[0]},\mathcal{Y}%
_{1}^{[0]})^{T}$ is a complex matrix solution of Lax pair~(\ref{lax pair})
with $\lambda (t)=\lambda _{1}(t)$, then $\Psi _{1}=(-\mathcal{Y}%
_{1}^{[0]\ast },\mathcal{H}_{1}^{[0]\ast })^{T}$ is a solution of Lax pair~(%
\ref{lax pair}) with $\lambda (t)=\lambda _{1}^{\ast }(t)$. The validity of
the above DT is established when the following conditions hold: $%
T[1]_{x}+T[1]U=U[1]T[1]$ and $T[1]_{t}+T[1]V=V[1]T[1]$, where $U[1]$ and $%
V[1]$ have the same form as $U$ and $V$, except that $Q[0]$ is replaced by $%
Q[1]$.

In the same way, taking $\Psi _{j}=(\mathcal{H}_{j}^{[0]},\mathcal{Y}%
_{j}^{[0]})^{T}$ ($j=1,2,3,\ldots $) to be complex matrix solutions of Lax
pair~(\ref{lax pair}) with $Q=Q[0]$ and $\lambda (t)=\lambda _{j}(t)$, and
iterating the first-order DT $N$ times, we construct the $N$-th-order DT for
system~(\ref{system}) as

\begin{widetext}
\begin{subequations} \label{DTN}
\begin{align}
  \Psi[N]=&T[N]T[N-1]\cdots T[1]\Psi,~~T[j]=\tau_j(\lambda)[\lambda I-H[j-1]\Lambda_jH[j-1]^{-1}],\\
  Q[N]=&Q[0]-\sum_{j=1}^N 2\text{i}(\lambda_j-\lambda_j^*)\left(\mathcal{Y}_j^{[j-1]}{\mathcal{H}_j^{[j-1]}}^{-1}
  +\mathcal{H}_j^{[j-1]}{\mathcal{Y}_j^{[j-1]}}^{-1}\right)^{-1},
\end{align}
\end{subequations}
\end{widetext}
where

\begin{equation}
\begin{aligned}
\Psi_j[j-1]=&(\mathcal{H}_j^{[j-1]},\mathcal{Y}_j^{[j-1]})^T\\
=&T[j-1]|_{\lambda=\lambda_j}T[j-2]|_{\lambda=\lambda_{j-1}}\cdots
T[1]|_{\lambda=\lambda_2}\Psi_j,\\
\tau_j(\lambda)=&[(\lambda(t)-\lambda_j(t))(\lambda(t)-\lambda_j^*(t))]^{-%
\frac{1}{2}},\\ H[j-1]=&\begin{pmatrix} \mathcal{H}_j^{[j-1]} &
-\mathcal{Y}_j^{[j-1]} \\ \mathcal{Y}_j^{[j-1]} & \mathcal{H}_j^{[j-1]}
\end{pmatrix}, \Lambda_j=\begin{pmatrix} \lambda_jI & O \\ O & \lambda_j^*I
\end{pmatrix}. \end{aligned}
\end{equation}

For the above $N$-th-order DT~(\ref{DTN}), the following remarks are
relevant. (1) As system~(\ref{system}) is nonautonomous with the temporal
modulation, the Lax pair~(\ref{lax pair}) is nonisospectral with the
time-dependent spectral parameter $\lambda (t)$, which further makes the DT
different from that for the spinor BEC system without the external
potential, $v_{\text{ext}}(x,t)=0$. (2) $\tau _{j}(\lambda )$ is essential
in the DT matrix to make the $t$-part of Lax pair~(\ref{lax pair}) to be
satisfied. (3) Because the occurrence of $\tau _{j}(\lambda )$ in the DT
matrix, the determinant representation of the above DT~(\ref{DTN}) can not
be expressed directly via Cramer's rule~\cite{Cramer rules}.

\section{Nonlinear dynamics of nonautonomous matter-wave solitons with the
spatiotemporal modulation}

In this section, utilizing the $N$-th-order DT~(\ref{DTN}) derived in the
previous Section, we construct nonautonomous matter-wave-soliton solutions
for system~(\ref{system}). To this end, taking the zero seed solution $%
Q[0]=O $, and substituting $Q[0]$ into Lax pair~(\ref{lax pair}), we derive
the matrix Jost function, $\Psi _{1}=(\mathcal{H}_{1}^{[0]},\mathcal{Y}%
_{1}^{[0]})^{T}$, as
\begin{equation}
\mathcal{H}_{1}^{[0]}=e^{-\theta }I,~~\mathcal{Y}_{1}^{[0]}=e^{\theta }\Pi
^{\ast },  \label{eigenfunction1}
\end{equation}%
where $\theta =\text{i}\left( \lambda _{1}(t)x+2\int \lambda
_{1}^{2}(t)\,dt\right) +\theta _{0}$,
\begin{equation}
\Pi =%
\begin{pmatrix}
a & b \\
b & c%
\end{pmatrix}%
,  \label{polarization matrix}
\end{equation}%
\begin{equation}  \label{lambda1}
\begin{aligned} \lambda _{1}(t)=& \left[ \xi _{1}-\frac{1}{2}\int \exp
\left( 2\int \Gamma (t)\,dt\right) \,\gamma (t)dt\right]\times\\ &\exp
\left( -2\int \Gamma (t)\,dt\right), \end{aligned}
\end{equation}
$\xi _{1}$, $a$, $b$ and $c$ are complex constants, and $\theta _{0}$ is a
real constant which can be used to adjust the initial position of the
solitons. We normalize the complex matrix $\Pi $ so that
\begin{equation}
\text{tr}\{\Pi ^{\dag }\cdot \Pi \}=|a|^{2}+2|b|^{2}+|c|^{2}=1.
\label{unity}
\end{equation}

Combining DT~(\ref{DT1}) and the above matrix Jost function~(\ref%
{eigenfunction1}), we obtain the one-soliton solutions
\begin{equation}
Q[1]=\frac{4\lambda _{1I}e^{2\theta ^{\ast }}}{\Delta }%
\begin{pmatrix}
a+c^{\ast }e^{4\theta _{R}}\text{det}\Pi & b-b^{\ast }e^{4\theta _{R}}\text{%
det}\Pi \\
b-b^{\ast }e^{4\theta _{R}}\text{det}\Pi & c+a^{\ast }e^{4\theta _{R}}\text{%
det}\Pi%
\end{pmatrix}
\label{one soliton}
\end{equation}%
where
\begin{equation}
\begin{aligned} \lambda _{1R}=& \left[ \xi _{1R}-\frac{1}{2}\int \exp \left(
2\int \Gamma (t)\,dt\right) \,\gamma (t)dt\right]\times\\ &\exp \left(
-2\int \Gamma (t)\,dt\right),\\ \lambda_{1I}=&~\xi_{1I}\exp \left( -2\int
\Gamma (t)\,dt\right),\\ \theta_I=&~\lambda_{1R}x+2\int
(\lambda_{1R}^2-\lambda_{1I}^2)\,dt,\\
\theta_R=&~\theta_0-\left(\lambda_{1I}x+4\int
\lambda_{1R}\lambda_{1I}\,dt\right),\\
\Delta=&~1+e^{4\theta_R}+e^{8\theta_R}|\text{det}\Pi|^2, \end{aligned}
\end{equation}%
the subscripts $R$ and $I$ representing the real and imaginary parts,
respectively. A more compact form of the one-soliton solutions~(\ref{one
soliton}) can be written as
\begin{equation}
Q[1]=4\lambda _{1I}\frac{e^{-2\theta _{R}}\left[ \Pi +(\sigma _{2}\Pi ^{\dag
}\sigma _{2})e^{4\theta _{R}}\text{det}\Pi \right] }{e^{-4\theta
_{R}}+1+e^{4\theta _{R}}|\text{det}\Pi |^{2}}e^{-2\text{i}\theta _{I}},
\label{one soliton1}
\end{equation}%
where $\sigma _{2}$ is the Pauli matrix. According to the above expressions,
we point out the relevance of each parameter for the solitons as follows: $%
\lambda _{1I}$, the imaginary part of the spectral parameter, is
proportional to the amplitude of soliton; its trajectory is determined by $%
\theta _{R}=0$; and $\Pi $ is the soliton's polarization matrix. Further, we
can conclude that the gain-loss strength $\Gamma (t)$ amplifies or
attenuates the soliton's amplitude, while $\gamma (t)$ has no effect on the
amplitude. The trajectory and velocity of the soliton are directly affected
by $\Gamma (t)$ and $\gamma (t)$. As for the polarization matrix $\Pi $, it
affects spin states of the the soliton. Accordingly, the one-soliton
solution may be divided in two types, \textit{viz}., those for which $\text{%
det}\Pi $ is zero or not, as first proposed in Ref.~\cite{Ieda2004}.

The local spin density of the one-soliton state is
\begin{equation}
\mathbf{f}(x,t)=\mathbf{\Phi ^{\dag }}\cdot \mathbf{f}\cdot \mathbf{\Phi }=%
\text{tr}\{Q^{\dag }\boldsymbol{\sigma }Q\}  \label{spin density}
\end{equation}%
\cite{Ieda2004}, where $\mathbf{f}=(f^{x},f^{y},f^{z})^{T}$, with $f^{x,y,z}$
being the $3\times 3$ spin-1 matrices and $\boldsymbol{\sigma }$ the vector
of the Pauli matrices. Then, the spin density is derived as
\begin{equation}
\mathbf{f}(x,t)=\frac{16\lambda _{1I}^{2}e^{4\theta _{R}}(1-e^{8\theta _{R}}|%
\text{det}\Pi |^{2})}{(1+e^{4\theta _{R}}+e^{8\theta _{R}}|\text{det}\Pi
|^{2})^{2}}\text{tr}\{\Pi ^{\dag }\boldsymbol{\sigma }\Pi \}.
\label{spin density1}
\end{equation}%
The explicit form for the density of the number of atoms can also be
obtained from Eq.~(\ref{conserved quantities2})
\begin{equation}
n(x,t)=\frac{16\lambda _{1I}^{2}e^{4\theta _{R}}[1+(4e^{4\theta
_{R}}+e^{8\theta _{R}})|\text{det}\Pi |^{2}]}{(1+e^{4\theta _{R}}+e^{8\theta
_{R}}|\text{det}\Pi |^{2})^{2}}.  \label{number density}
\end{equation}

The above expressions for the nonautonomous one-soliton solution and its
spin and number-of-atoms densities are similar to those for the exact
solutions for autonomous solitons derived in Ref.~\cite{Ieda2004}. However,
due to the influence of the external potential, characteristics of the
nonautonomous solitons for system~(\ref{system}), such as the amplitude,
velocity, width, etc., are significantly different from their counterparts
in the case of the autonomous solitons. Next, we analyze the dynamics of the
nonautonomous solitons under the action of various external potentials.

\subsection{Time-independent external potential and gain-loss distribution}

First, we consider the simplest case of the external potential with a
constant gain-loss coefficient, $\Gamma (t)\equiv \Omega $, where $\Omega $
is the real constant. In this case, integrability condition (\ref{condition}%
) determines the time-independent HO potential in Eq. (\ref{UUU}) with
strength $U_{p}=\Omega ^{2}$.

\subsubsection{The nonautonomous ferromagnetic soliton}

To separately analyze the influence of the gain-loss coefficient $\Omega $
and linear-potential's coefficient $\gamma (t)$ (see Eq. (\ref{UUU})) on the
dynamics of solitons, we first set $\Omega \neq 0$ and $\gamma (t)=0$. Then
the external potential (\ref{vext}) is $v_{\text{ext}}(x,t)=\Omega ^{2}x^{2}+%
\text{i}\Omega $ and we set $\text{det}\Pi =0$. Then, the one-soliton
solution~(\ref{one soliton1}) reduces
\begin{equation}
Q[1]=2\lambda _{1I}\,\text{sech}(2\theta _{R})e^{-2\text{i}\theta _{I}}\Pi ,
\label{fm soliton1}
\end{equation}%
where
\begin{equation}
\begin{aligned} \lambda_{1I}=&~\xi_{1I}\,e^{-2\Omega
t},~~\lambda_{1R}=\xi_{1R}\,e^{-2\Omega t},\\ \theta_I=&~\lambda_{1R}x+2\int
(\lambda_{1R}^2-\lambda_{1I}^2)\,dt,\\
\theta_R=&~\theta_0-\left(\lambda_{1I}x+4\int
\lambda_{1R}\lambda_{1I}\,dt\right). \end{aligned}
\end{equation}%
It can be seen that the three components $\{\phi _{+1},\phi _{0},\phi
_{-1}\} $ share the same bell-like shape. Solutions~(\ref{fm soliton1})
yields the amplitude of the soliton as $A=2|\lambda _{1I}|(|a|,|b|,|c|)^{T}$
for the three components $\{\phi _{+1},\phi _{0},\phi _{-1}\}$, velocity $%
v=-2\xi _{1R}\,e^{-2\Omega t}+2\Omega \theta _{0}\,e^{2\Omega t}/\xi _{1I}$,
and the width which is proportional to $e^{2\Omega t}/|\xi _{1I}|$. These
results indicate that the amplitude, velocity and width vary with time under
the action of the gain-loss term $\sim \Omega $.

\begin{figure}[th]
\includegraphics[scale=0.55]{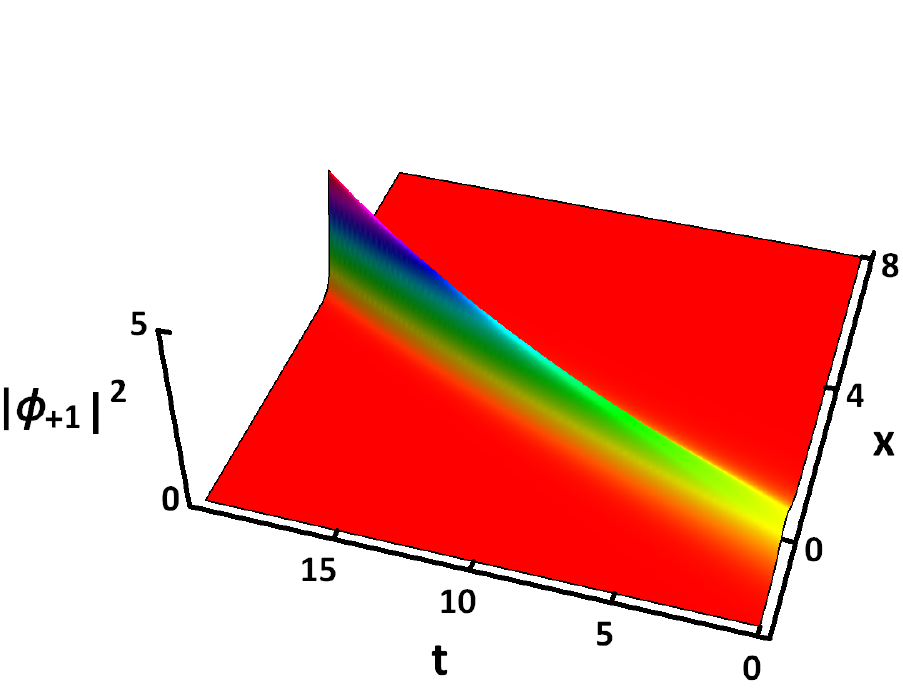}\hspace{2mm} %
\includegraphics[scale=0.55]{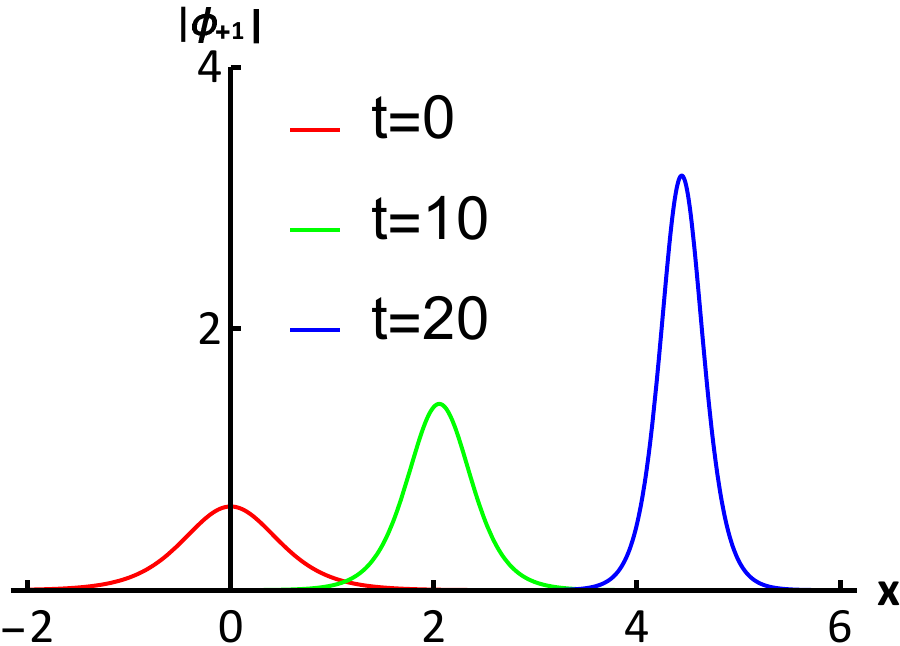}\hspace{2mm} %
\includegraphics[scale=0.55]{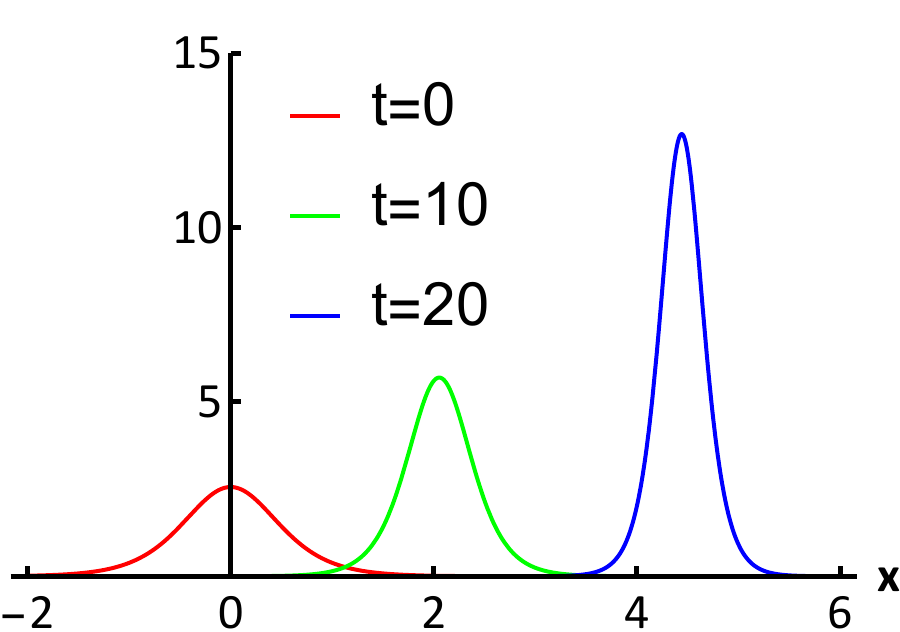}
\caption{(Top) The 3D plot of the nonautonomous FS soliton of component $%
\protect\phi _{+1}$. (Middle) The profile of the soliton at different times.
(Bottom) The atomic number density $n(x,t)$ at different times. The external
potential is $v_{\text{ext}}(x,t)=\Omega ^{2}x^{2}+\text{i}\Omega $ and
parameters are $\protect\xi _{1}=-0.05+0.8\text{i}$, $\text{det}\Pi =0$, $%
\Omega =-0.02$, $\protect\theta _{0}=-2$.}
\label{fig1}
\end{figure}

The atomic-number density for this solution is $n(x,t)=4\lambda _{1I}^{2}\,%
\text{sech}^{2}(2\theta _{R})$, and respective total number of atoms is $%
N_{T}=4\xi _{1I}\,\exp \left( -2\Omega t\right) \equiv N_{T0}\,\exp \left(
-2\Omega t\right) $, where $N_{T0}=4\xi _{1I}$ is the initial value of the
total number. The spin density of the solution is $\mathbf{f}(x,t)=4\lambda
_{1I}^{2}\,\text{sech}^{2}(2\theta _{R})\text{tr}\{\Pi ^{\dag }\boldsymbol{%
\sigma }\Pi \}=n(x,t)\text{tr}\{\Pi ^{\dag }\boldsymbol{\sigma }\Pi \}$.
Then, the total spin can be calculated as $\mathbf{F}_{T}=4\xi _{1I}\,\exp
\left( -2\Omega t\right) \text{tr}\{\Pi ^{\dag }\boldsymbol{\sigma }\Pi \}$,
and $|\mathbf{F}_{T}|=N_{T}$ for $\text{det}\Pi =0$. Note that the spin
density has the same profile as the atomic-number density. With a nonzero
total spin, this solution is referred to as the ferromagnetic-state (FS)
soliton~\cite{Ieda2004}. As it moves with varying amplitude and velocity, it
is also classified as a nonautonomous soliton~\cite{Serkin2007}.

In Fig.~\ref{fig1}, we show the density profile of the nonautonomous FS
soliton for component $\phi _{+1}$ (the other two components $\phi _{0}$ and
$\phi _{-1}$ have similar shapes) and the atomic-number density
distribution, $n(x,t)$. It is seen that the width of the soliton decreases
and its amplitude increases in the course of the evolution at $\Omega <0$
(the case of the gain), as seen in Fig.~\ref{fig1}. On the contrary, if $%
\Omega >0$ (the loss), the width of the soliton increases and amplitude
decreases.

\subsubsection{The nonautonomous polar soliton}

In this case, we set $\text{det}\Pi \neq 0$ and take the same external
potential as above, $v_{\text{ext}}(x,t)=\Omega ^{2}x^{2}+\text{i}\Omega $.
The expression for the spin density~(\ref{spin density1}) shows that the
total spin is always nonzero, i.e., $|\mathbf{F}_{T}|\neq 0$, unless $\Omega
=0$ or $2\,|\text{det}\Pi |=1$. Because of the presence of $\lambda
_{1I}=\xi _{1I}\,e^{-2\Omega t}$ in spin density~(\ref{spin density1}),
which is a time-dependent function, the total spin cannot be zero, which is
strongly different from the solution of the integrable spinor BEC system
with constant coefficients \cite{Ieda2004}. There are two special cases that
make the total spin to be zero, \textit{viz}., $\Omega =0$ or $2\,|\text{det}%
\Pi |=1$. In the case of $\Omega =0$, $\lambda _{1I}$ is time-independent,
and a coordinate transformation makes it possible to make the spin density
function $\mathbf{f}(x,t)$ an odd function, so the total spin $|\mathbf{F}%
_{T}|$ is zero. In latter case, the constrain $2\,|\text{det}\Pi |=1$, along
with the normalization condition $\text{tr}\{\Pi ^{\dag }\cdot \Pi \}=1$,
leads to a result $\text{tr}\{\Pi ^{\dag }\boldsymbol{\sigma }\Pi \}\equiv
(0,0,0)^{T}$. From the expression for the spin density~(\ref{spin density1}%
), it then follows that the spin density vanishes everywhere, i.e., $\mathbf{%
f}(x,t)\equiv (0,0,0)^{T}$. Solitons in this state hold the local symmetry
of the polar state. Therefore, the solutions given by Eq. (\ref{one soliton1}%
) with $2\,|\text{det}\Pi |=1$ are referred to the polar-state (PS)
solitons, which feature the bell-shaped profile, similar to the FS solitons.

\begin{figure}[th]
\includegraphics[scale=0.45]{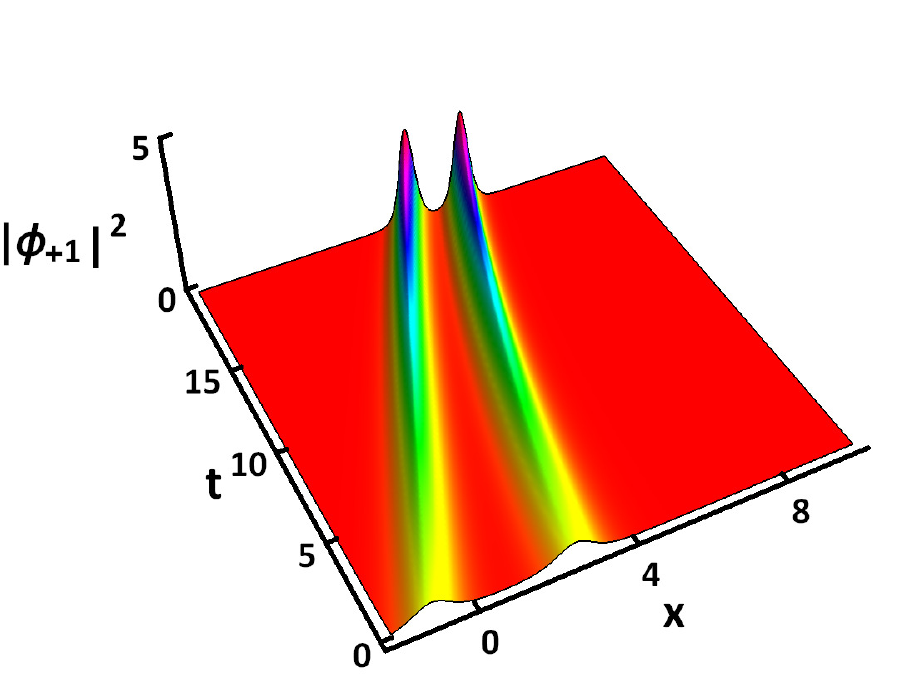}\hspace{1mm} %
\includegraphics[scale=0.45]{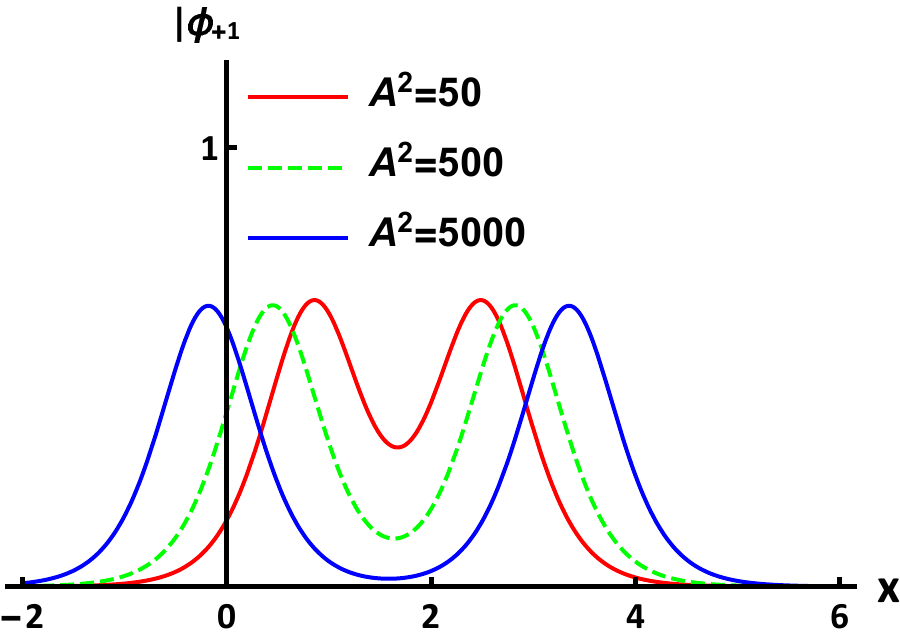} {%
\center{\footnotesize
($a$)\hspace{3cm}($b$)}}\newline
\caption{(a) The 3D plot of the nonautonomous PS soliton in component $%
\protect\phi _{+1}$ with $A^{2}=5000$. (b) The profile of the soliton at $%
t=0 $ with a different polarization matrix $\Pi $. The external potential is
$v_{\text{ext}}(x,t)=\Omega ^{2}x^{2}+\text{i}\Omega $, and the parameters
are $\protect\xi _{1}=-0.05+0.8\text{i}$ and $\Omega =-0.02$.}
\label{fig2}
\end{figure}

When polarization matrix is constrained by $2\,|\text{det}\Pi |<1$, the spin
density is nonzero. In this case, the single peak of the density split in
two for the three components $\{\phi _{+1},\phi _{0},\phi _{-1}\}$, which is
different from the FS soliton with $\text{det}\Pi =0$ and PS soliton with $%
2\,|\text{det}\Pi |=1$. Further, as $|\text{det}\Pi |$ decreases, the twin
peaks gradually separate, and behave like two nonautonomous FS solitons.
Such solutions are referred to as nonautonomous split solitons. According to
solutions~(\ref{one soliton1}), three components $\{\phi _{+1},\phi
_{0},\phi _{-1}\}$ have similar profiles. In Fig.~\ref{fig2} we display the
density profile of the nonautonomous PS soliton in component $\phi _{+1}$
with $A^{2}=5000$, where $A^{-1}=2|\text{det}\Pi |$. As said above, the twin
peaks of the PS soliton gradually separate with the increase of $A$, as
illustrated in Fig.~\ref{fig2}(b). Due to the effect of the gain-loss term,
the twin peaks of the split soliton slowly approach, rather than staying
parallel, which is different from the case of autonomous integrable system~%
\cite{Ieda2004}. Parameter $\theta _{0}$ has been used to adjust the
positions of the twin peaks with different $A$ in Fig.~\ref{fig2}(b).

\subsubsection{Numerical simulations of the evolution of the nonautonomous
matter-wave solitons}

\begin{figure}[th]
\includegraphics[scale=0.25]{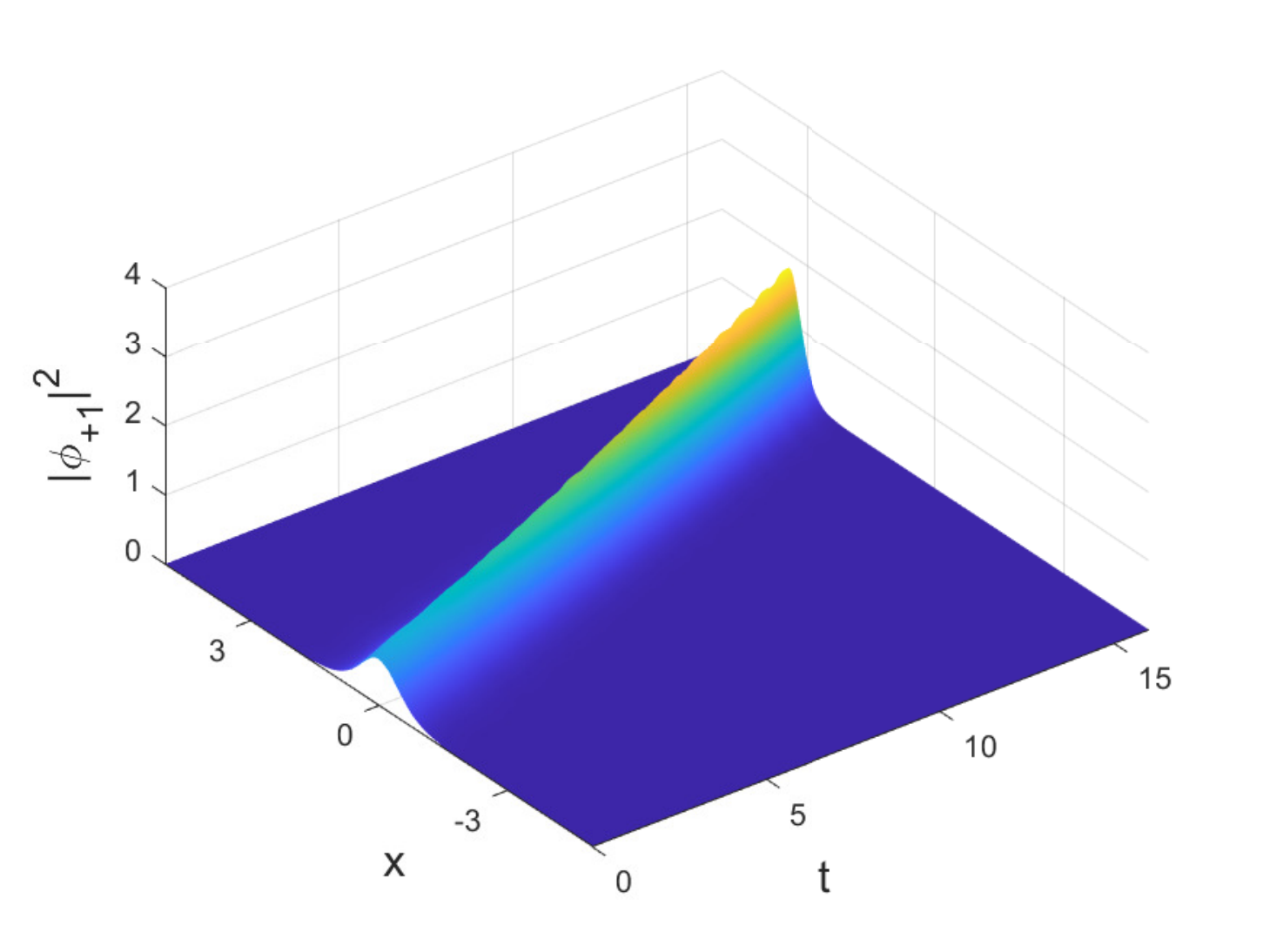}\hspace{1mm} %
\includegraphics[scale=0.25]{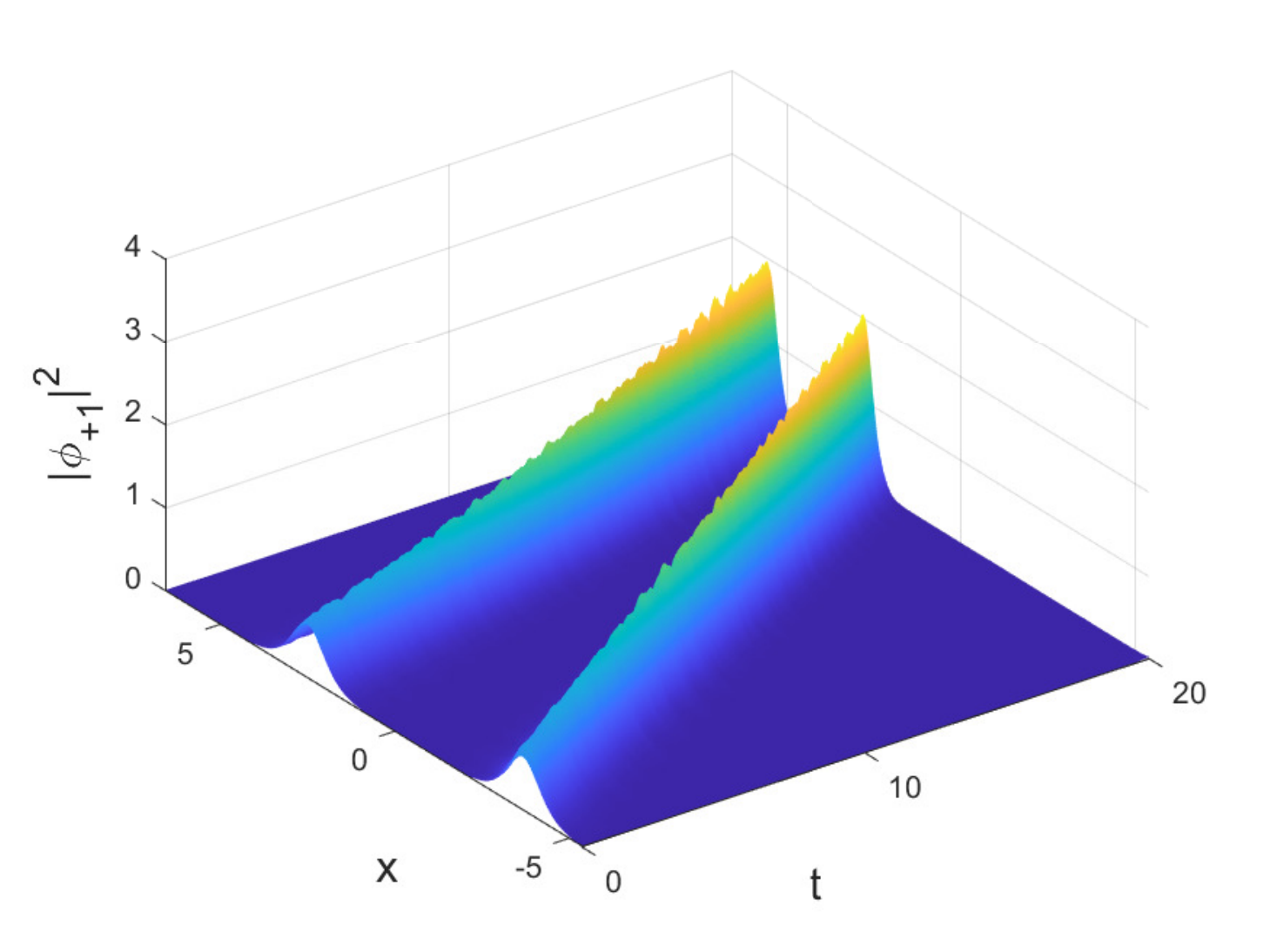} {%
\center{\footnotesize ($a$)
$A=0$\hspace{2.5cm}($b$) $A^2=5000$}}\newline
\caption{Numerical simulations of the FS and PS solitons. Inputs for the
numerical simulations are taken as the one-soliton solution (\protect\ref%
{one soliton1}) with a random noise level added to it at the $2\%$ level.
The other parameters are taken the same as in Figs.~\protect\ref{fig1} and~%
\protect\ref{fig2}.}
\label{fig151}
\end{figure}

Next, we use numerical simulations to analyze the stability of the
nonautonomous matter-wave solitons. The split-step Fourier method has been
applied to analyze the stability in the case of the time-independent
external potential, $v_{\text{ext}}=\Omega ^{2}x^{2}+\text{i}\Omega $, and
constant gain-loss coefficient, $\Gamma (t)\equiv \Omega $. In the
simulations, random noise is added to the input taken as the one-soliton
solution~(\ref{one soliton1}). The stability of the nonautonomous
matter-wave solitons is verified for both the ferromagnetic and polar
states, as shown in Fig.~\ref{fig151}. It is seen that both the FS and PS
solitons are indeed dynamically stable in the presence of the random noise.

\subsubsection{Nonautonomous solitons without the gain-loss term}

In this case, we set $\Omega =0$ with $\gamma (t)\neq 0$ to analyze the
effects of the external potential (\ref{vext}), which reduces to $v_{\text{%
ext}}(x,t)=\gamma (t)x$. In this condition, the velocity of the soliton is $%
v=2\int \gamma (t)\,dt-4\xi _{1R}$, its width is inversely proportional to $%
|\xi _{1I}|$, and the amplitude of the soliton is $A=2|\xi
_{1I}|(|a|,|b|,|c|)^{T}$ for the three components $\{\phi _{+1},\phi
_{0},\phi _{-1}\}$.

\begin{figure}[th]
\includegraphics[scale=0.45]{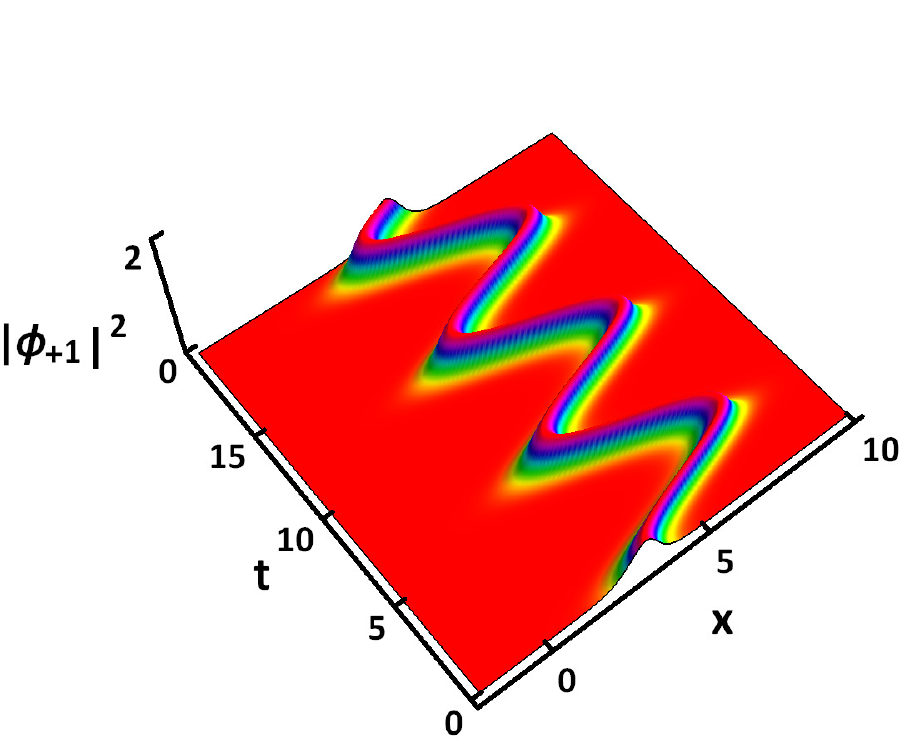}\hspace{1mm} %
\includegraphics[scale=0.45]{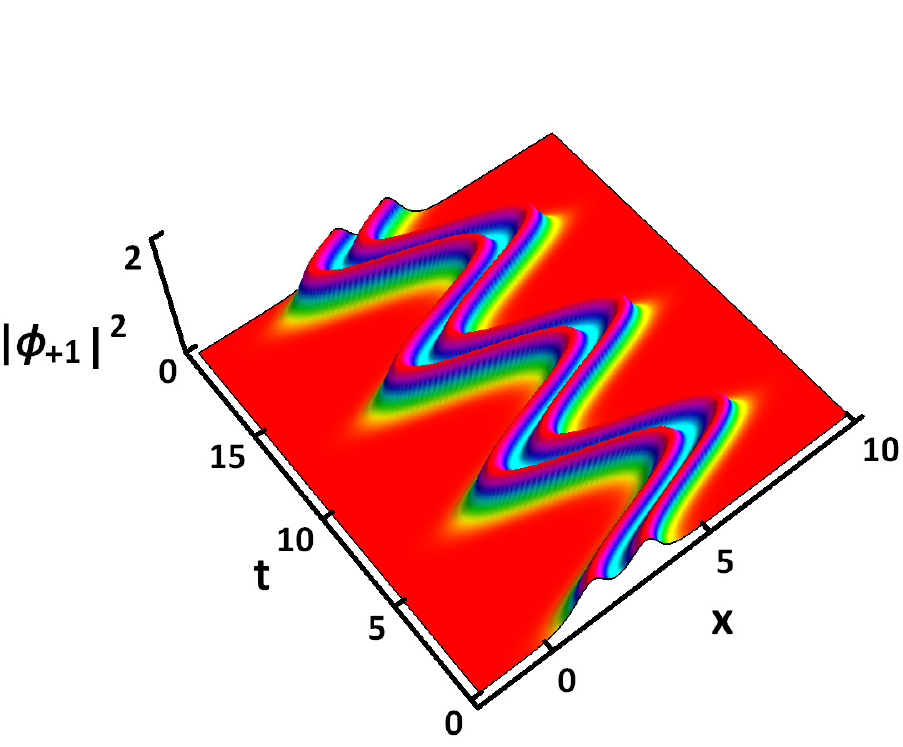} {%
\center{\footnotesize
($a$)\hspace{4cm}($b$)}}\newline
\caption{(a) The 3D plot of the nonautonomous FS soliton in component $%
\protect\phi _{+1}$ with $A=0$. (b) The 3D plot of the nonautonomous split
soliton in component $\protect\phi _{+1}$ with $A^{2}=50$. The external
potential is $v_{\text{ext}}(x,t)=(\cos t)x$ and the parameters are $\protect%
\xi _{1}=0.8\text{i}$ and $\protect\theta _{0}=4$.}
\label{fig3}
\end{figure}

It is seen that $\gamma (t)$ has no effect on the amplitude of the soliton,
but notably affect the velocity of the soliton. In Fig.~\ref{fig3}, by
choosing a $\gamma (t)=\cos (t)$, snakelike FS and PS solitons are obtained
with a constant amplitude and periodically varying velocity.

As the gain-loss term is absent, the total number of atoms $N_{T}=4\xi
_{1I}=N_{T0}$ remains constant, whether $\text{det}\Pi =0$ or not, i.e., the
total number of atoms in system~(\ref{system}) is conserved even in the
presence of the time-dependent coefficient $\gamma (t)$ in the linear
potential. As for the spin density and total spin, it is found that $|%
\mathbf{F}_{T}|=N_{T}$ for $\text{det}\Pi =0$. When $2|\text{det}\Pi |=1$,
spin density $\mathbf{f}(x,t)$ vanishes everywhere and, naturally, the total
spin vanishes too, $|\mathbf{F}_{T}|=0$. On the other hand, for $2|\text{det}%
\Pi |<1$ the spin density $\mathbf{f}(x,t)$ remains nonzero, even though the
total spin is again zero, $|\mathbf{F}_{T}|=0$.

\subsection{The time-dependent external potential and gain-loss distribution}

Next, we consider the case when the HO potential $U_{p}(t)$ in Eq. (\ref{UUU}%
) is time-dependent and can be attractive or expulsive (inverted HO).
Various potential functions $U_{p}(t)$ and gain-loss coefficients $\Gamma
(t) $ can be used to generate different nonautonomous matter-wave solitons.
We here choose certain physically relevant forms of the external potential
to investigate the evolution of several kinds of nonautonomous solitons. In
this case, we set $\gamma (t)=0$ in Eq. (\ref{UUU}).

\subsubsection{Nonautonomous solitons with a step-wise time-modulated
gain-loss coefficient}

To amplify a soliton with a small amplitude into one with an appropriate
amplitude, the following step-wise gain-loss coefficient can be used:
\begin{equation}
\Gamma (t)=\rho \lbrack 1+\tanh (2\rho t)],  \label{step-wise gain}
\end{equation}%
where $\rho $ determined the steepness of the step and $\delta $ is the
initial phase of the step. Then, according to the integrability condition (%
\ref{condition}), the time-dependent external potential $U_{p}(t)$ (\ref{UUU}%
) is
\begin{equation}
U_{p}(t)\equiv \frac{1}{2}\Omega ^{2}(t)=2\rho ^{2}[1+\tanh (2\rho t)].
\label{step-wise Up}
\end{equation}

\begin{figure}[th]
\includegraphics[scale=0.7]{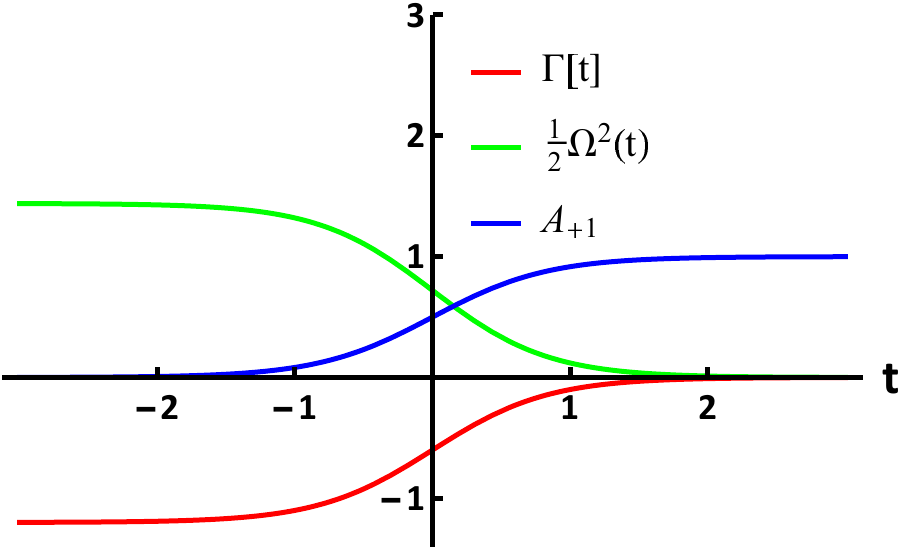}
\caption{The kink-like shape of the gain-loss coefficient $\Gamma (t)$ (the
red line), HO potential strength $(1/2)\Omega ^{2}(t)$ (the green line), and
amplitude $A_{+1}$ for component $\protect\phi _{+1}$ (the blue line). The
parameters are $\protect\rho =-0.6$, $\protect\delta =0$, $a=0.5$ and $%
\protect\xi _{1I}=-0.5$.}
\label{fig4}
\end{figure}
In this case, solution~(\ref{one soliton1}), yields the amplitude of the
nonautonomous soliton for the three components, $A=2|\lambda
_{1I}|(|a|,|b|,|c|)^{T}$ where
\begin{equation}
\lambda _{1I}=\xi _{1I}\,\exp \left( -2\int \Gamma (t)\,dt\right) =\xi
_{1I}\,e^{\delta }[1-\tanh (2\rho t)].  \label{step-wise amplitude}
\end{equation}

In Fig.~\ref{fig4}, we show the kink-like shape of the gain-loss coefficient
$\Gamma (t)$, HO potential strength $(1/2)\Omega ^{2}(t)$, and amplitude $%
A_{+1}$ of component $\phi _{+1}$. Obviously, $\rho <0$ and $\rho >0$ in Eq.
(\ref{step-wise gain}) correspond to $\Gamma (t)\leq 0$ and $\Gamma (t)\geq
0 $ (the gain and loss), respectively. Accordingly, the parabolic potential
and amplitude of the soliton are always non-negative, with the former and
latter step-wise decreasing to zero or increasing from zero to a finite
value, respectively.

\begin{figure}[th]
\includegraphics[scale=0.45]{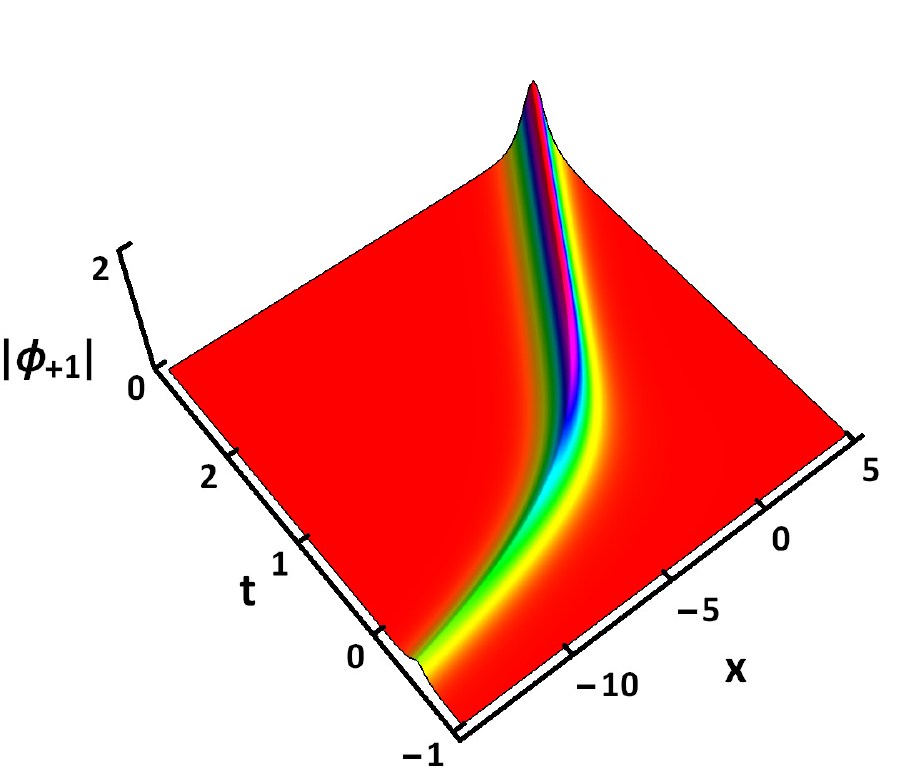}\hspace{1mm} %
\includegraphics[scale=0.45]{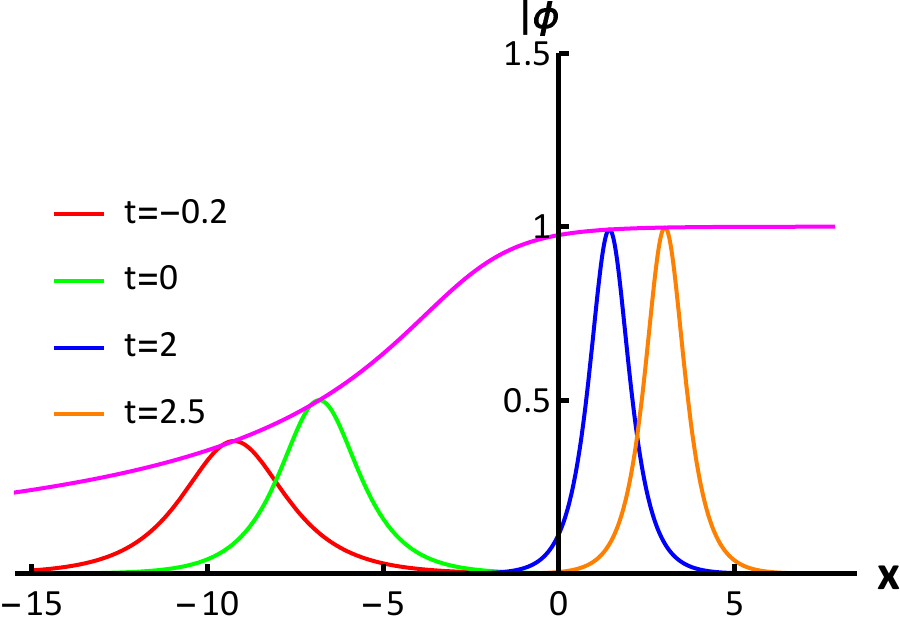} {%
\center{\footnotesize
($a_1$)\hspace{4cm}($a_2$)}}\newline
\includegraphics[scale=0.45]{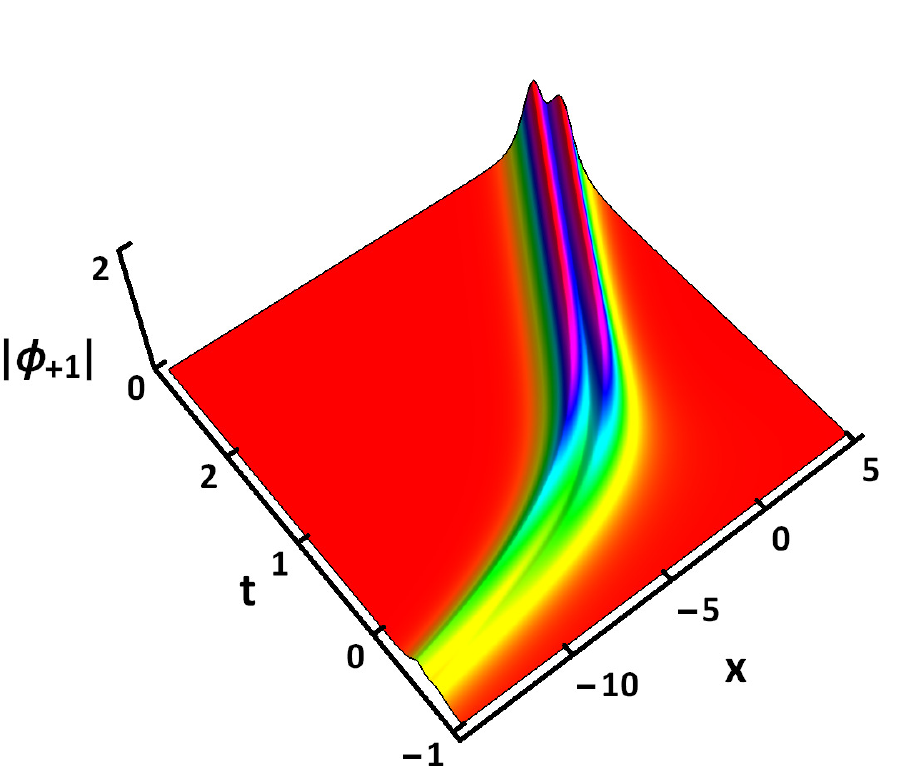}\hspace{1mm} %
\includegraphics[scale=0.45]{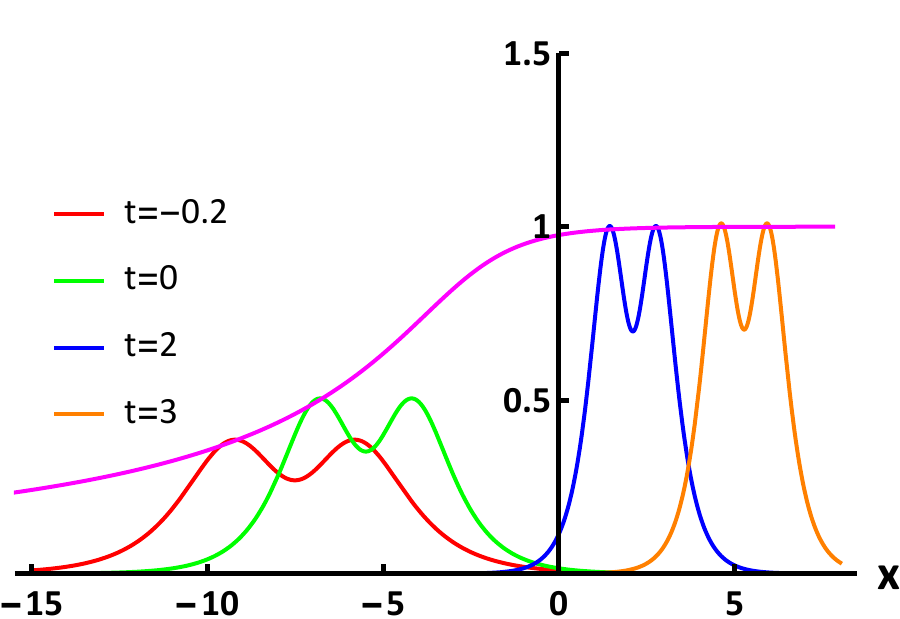} {%
\center{\footnotesize
($b_1$)\hspace{4cm}($b_2$)}}\newline
\caption{Density profiles of ($a_{1}$) the nonautonomous FS soliton with $%
A=0 $ and ($b_{1}$) nonautonomous split soliton with $A^{2}=50$ in component
$\protect\phi _{+1}$. ($a_{2}$) and ($b_{2}$) are soliton profiles at
different times and peak values (the magenta line) of the soliton fields.
The parameters are $\protect\rho =-0.5$, $\protect\delta =0$, $\protect\xi %
=-0.4-0.5\text{i, }$and $\protect\theta _{0}=5$.}
\label{fig5}
\end{figure}

Density profiles of the two kinds of nonautonomous solitons, FS and PS ones,
are displayed in Fig.~\ref{fig5}. Under the step-wise gain effect ($\Gamma
(t)<0$), the nonautonomous solitons are amplified into states with finite
constant amplitudes both for the FS and PS solitons. Unlike the previous
solitons whose amplitudes increase exponentially with time, the amplitudes
of these solitons in Fig.~\ref{fig5} grow to a finite value and remain
unchanged, as shown by the magenta curves in Figs.~\ref{fig5}($a_{2}$) and ($%
b_{2}$). To explain this, we find from Eqs.~(\ref{step-wise gain}), (\ref%
{step-wise amplitude}) and Fig.~\ref{fig4} that, when the amplitude of the
soliton attains the maximum, the gain coefficient $\Gamma (t)$ falls to
zero. Furthermore, we find that the PS soliton is asymmetric at the
amplitude amplification stage, but when the amplitude reaches the maximum,
the double-hump soliton becomes symmetric.

In this case, we find the atom-number density $n(x,t)=4\lambda _{1I}^{2}%
\text{sech}^{2}(2\theta _{R})$ and total number of atoms $N_{T}=4\xi
_{1I}e^{\delta }[1-\tanh (2\rho t)]$ when $\text{det}\Pi =0$, where $\lambda
_{1I}$ is given by Eq.~(\ref{step-wise amplitude}). It is found that the
time dependence of the total number of atoms $N_{T}$ is also kink-like,
similar to the amplitude of soliton shown in Fig.~\ref{fig4}. When $2|\text{%
det}\Pi |=1$, the number density is derived as $n(x,t)=8\lambda _{1I}^{2}$%
sech$^{2}(2\theta _{R}-(1/2)\ln 2)$, and the total number of atoms is $%
N_{T}=8\xi _{1I}e^{\delta }[1-\tanh (2\rho t)]$, which is two times larger
than in the case of $\text{det}\Pi =0$. As for the spin density and total
spin under the condition $\text{det}\Pi =0$, it is found that the spin
density is $\mathbf{f}(x,t)=4\lambda _{1I}^{2}\text{sech}^{2}(2\theta _{R})%
\text{tr}\{\Pi ^{\dag }\boldsymbol{\sigma }\Pi \}$, and the total spin is $%
\mathbf{F}_{T}=4\xi _{1I}e^{\delta }[1-\tanh (2\rho t)]\text{tr}\{\Pi ^{\dag
}\boldsymbol{\sigma }\Pi \}$ with $|\mathbf{F}_{T}|=4\xi _{1I}e^{\delta
}[1-\tanh (2\rho t)]$. We point out that\ both the total number of atoms and
total spin exhibit the same kink-like shape time dependence as the amplitude
of the soliton. When $2|\text{det}\Pi |=1$, we find that the spin density
vanishes everywhere, i.e., $\mathbf{f}(x,t)\equiv (0,0,0)^{T}$, which is a
characteristic of the polar state. When the polarization matrix is
constrained by $2|\text{det}\Pi |<1$, the spin density $\mathbf{f}(x,t)$ is
not zero but the total spin vanishes, $|\mathbf{F}_{T}|=0$, as shown by the
direct calculation.

\subsubsection{Nonautonomous solitons with a double-modulated periodic
potential}

Above, we have analyzed the effects of the periodic modulation of the
coefficient $\gamma (t)$ in front of the linear potential in Eq. (\ref{UUU})
on the soliton speed. Here we modulate the amplitude of the soliton by
applying the time-periodic modulation to the gain-loss coefficient $\Gamma
(t)$ and HO-potential strength $U_{p}(t)$. Namely, we set $\gamma (t)=0$ and
\begin{equation}
\Gamma (t)=\Omega _{0}\sin (\omega t),  \label{periodic gain}
\end{equation}%
where $\Omega _{0}$ and $\omega $ are the intensity and frequency of the
periodic modulation. Then, according to the integrability condition (\ref%
{condition}), we obtain the following double-harmonic modulation format,
\begin{equation}
U_{p}(t)\equiv \frac{1}{2}\Omega ^{2}(t)=\frac{1}{2}\Omega _{0}\{\Omega
_{0}[1-\cos (2\omega t)]+\omega \cos (\omega t)\}.  \label{periodic Up}
\end{equation}

\begin{figure}[th]
\includegraphics[scale=0.7]{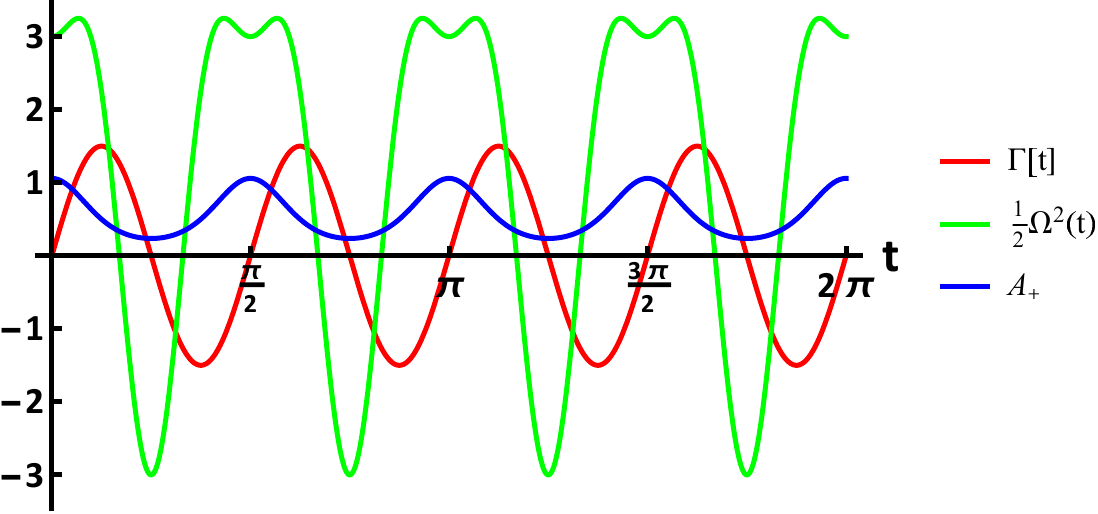}
\caption{The periodically modulated gain-loss coefficient $\Gamma (t)$ (the
red line), HO-potential strength$\ (1/2)\Omega ^{2}(t)$ (the green line) and
amplitude $A_{+1}$ for component $\protect\phi _{+1}$ (the blue line). The
parameters are $\protect\omega =4$, $\Omega _{0}=1.5$, $a=0.5$, and $\protect%
\xi _{1I}=0.5$.}
\label{fig6}
\end{figure}

\begin{figure}[th]
\includegraphics[scale=0.7]{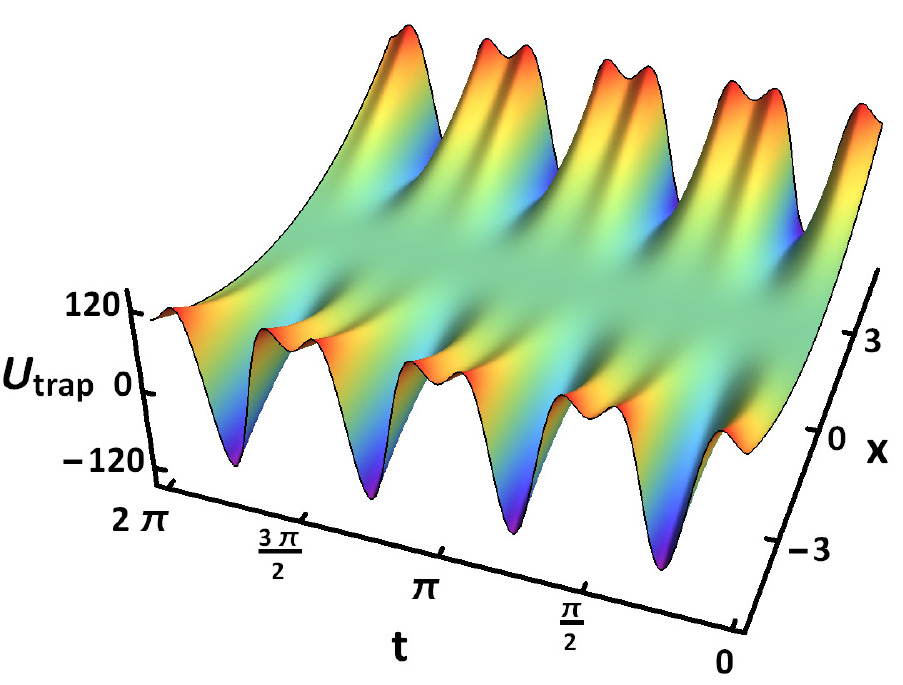}
\caption{Periodic sign-changing oscillations of the trapping HO potential $%
U_{\text{trap}}(x,t)=(1/2)\Omega ^{2}(t)x^{2}$, cf. a similar time-modulated
potential which may maintain stable 2D solitons \protect\cite{Sandy}. The
parameters are $\protect\omega =4$ and $\Omega _{0}=1.5$.}
\label{fig7}
\end{figure}

Under the action of this modulation, the amplitude of the nonautonomous
soliton is $A=2|\lambda _{1I}|(|a|,|b|,|c|)^{T}$ for the three components,
where
\begin{equation}
\lambda _{1I}=\xi _{1I}\,\exp \left( -2\int \Gamma (t)\,dt\right) =\xi
_{1I}\,\exp \left( 2\frac{\Omega _{0}}{\omega }\cos (\omega t)\right) .
\label{periodic amplitude}
\end{equation}%
In Eqs.~(\ref{periodic gain})-(\ref{periodic amplitude}), the gain-loss
coefficient $\Gamma (t)$ periodically varies between $-\Omega _{0}$ to $%
+\Omega _{0}$ with period $2\pi /\omega $, while $U_{p}(t)$ attains the
minimum $-\omega \Omega _{0}/2$ at $t=(2k+1)\pi /\omega $ and maximum $%
\left( \omega /4\right) ^{2}+\Omega _{0}^{2}$ at $t=\left[ \arctan \left(
\omega /\Omega _{0},\pm \sqrt{16\Omega _{0}^{2}-\omega ^{2}}/\Omega
_{0}\right) +2k\pi \right] /\omega $, where $k$ is an integer. The
fluctuation range of the amplitude of the nonautonomous soliton is $2\xi
_{1I}|a|\left[ \exp \left( -2\Omega _{0}/\omega \right) ,\exp \left(
-2\Omega _{0}/\omega \right) \right] $ for component $\phi _{+1}$, with
period $2\pi /\omega $. We display the gain-loss coefficient $\Gamma (t)$,
HO-potential strength $(1/2)\Omega ^{2}(t)$, and amplitude $A_{+1}$ in Fig.~%
\ref{fig6}. It is seen that the $\Gamma (t)$ and $(1/2)\Omega ^{2}(t)$
perform sign-changing oscillations. We also show the periodic sign-changing
oscillations of the trapping HO potential, $U_{\text{trap}}(x,t)=(1/2)\Omega
^{2}(t)x^{2}$ in Fig.~\ref{fig7}. Nonautonomous solitons, including the FS
and PS, ones are stably maintained by the attractive-expulsive sign-changing
potential $U_{\text{trap}}(x,t)$, cf. a similar result recently reported for
2D solitons in Ref. \cite{Sandy}.

\begin{figure}[th]
\includegraphics[scale=0.45]{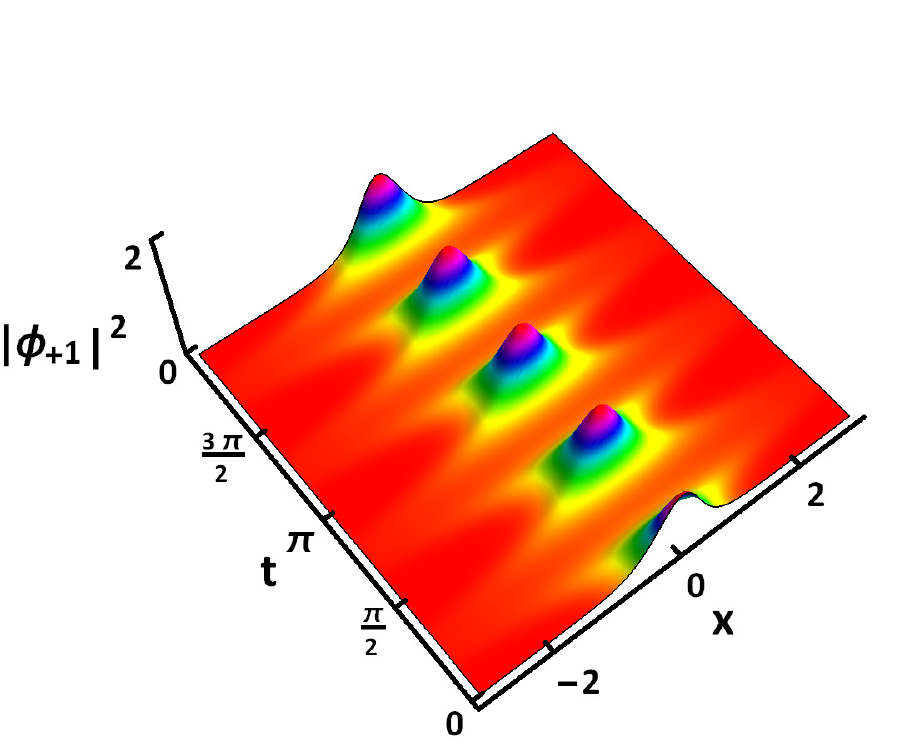}\hspace{1mm} %
\includegraphics[scale=0.45]{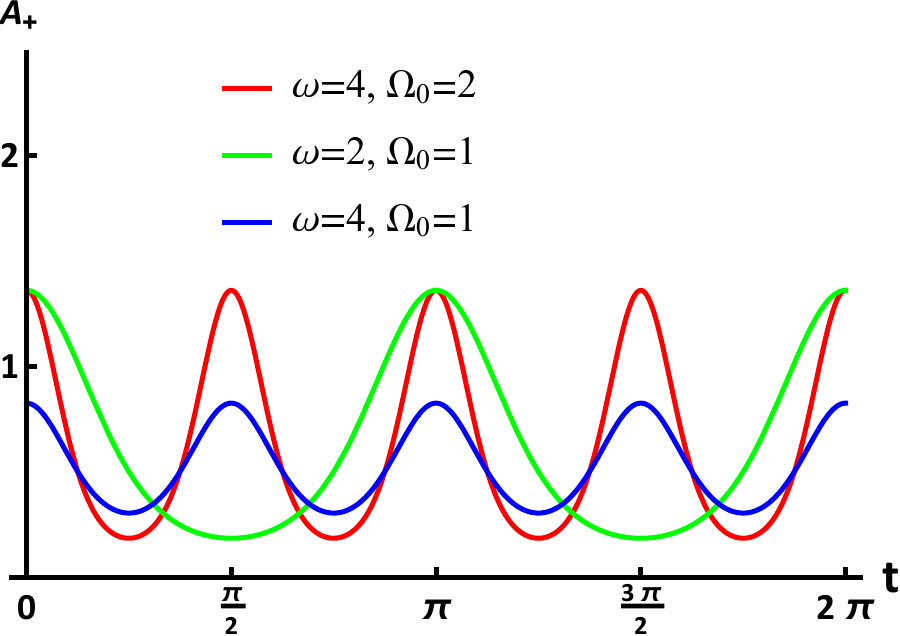} {%
\center{\footnotesize ($a_1$) $A=0
$\hspace{3cm}($a_2$)}}\newline
\includegraphics[scale=0.45]{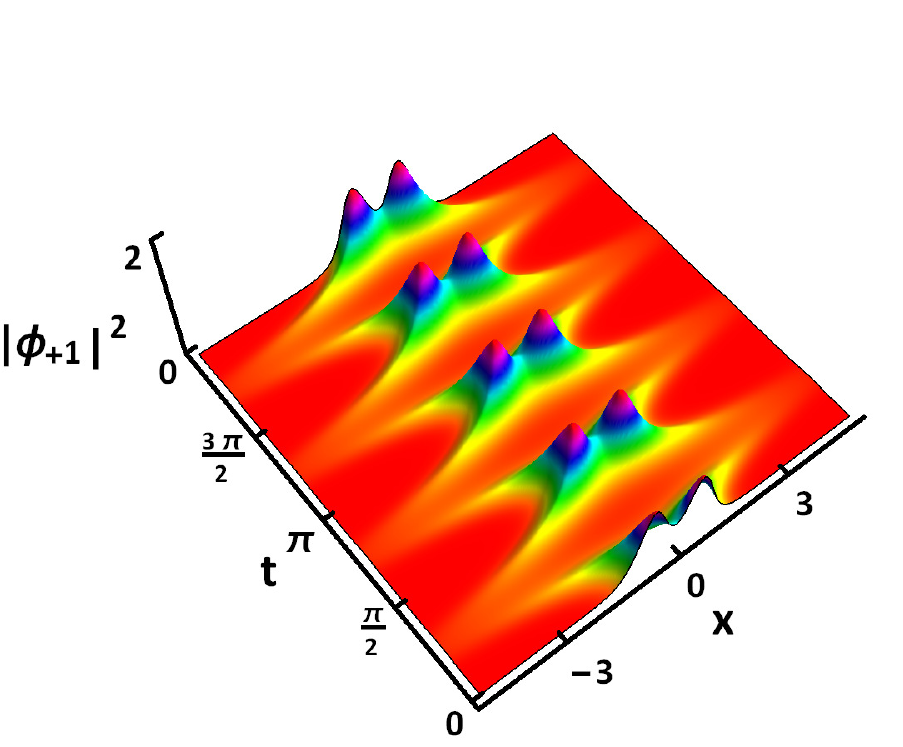}\hspace{1mm} %
\includegraphics[scale=0.45]{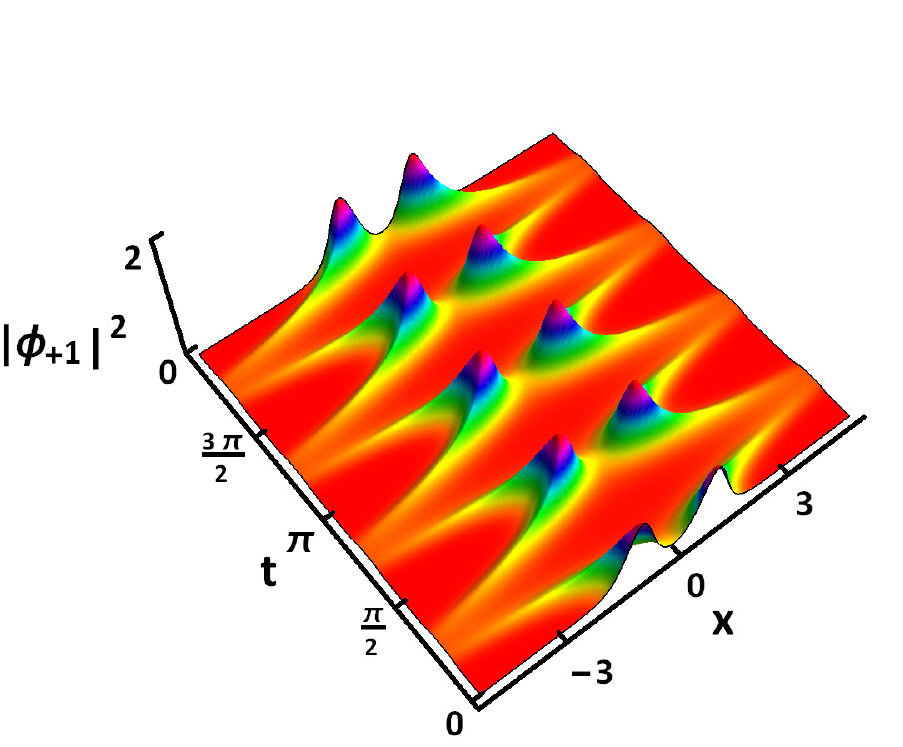} {%
\center{\footnotesize ($b_1$)
$A^2=50$\hspace{2cm}($b_2$) $A^2=1000$}}\newline
\caption{Density profiles of ($a_{1}$) the nonautonomous FS soliton with $%
\protect\theta _{0}=0$, and ($b_{1}$) nonautonomous PS soliton with $\protect%
\theta _{0}=0.6$ in component $\protect\phi _{+1}$. ($a_{2}$) The effect of
frequency $\protect\omega $ and ratio $\Omega _{0}/\protect\omega $ on the
breathing dynamics. ($b_{2}$) Nonautonomous PS soliton with $\protect\theta %
_{0}=1$. The parameters are $\protect\omega =4$, $\Omega _{0}=1.5$ and $%
\protect\xi _{1}=0.5\text{i}$. }
\label{fig8}
\end{figure}

Two kinds of breather solitons supported by the periodic modulation of the
potential are shown in Fig.~\ref{fig8}. Analysis of effects of frequency $%
\omega $ and scaled amplitude $\Omega _{0}/\omega $ of the modulation (see
Eqs. (\ref{periodic gain}) and (\ref{periodic Up}) demonstrates that the
breathing amplitude is proportion to $\Omega _{0}/\omega $, as shown in Fig.~%
\ref{fig8}($a_{2}$). Similarly, when $A\neq 0$, i.e., the polarization
matrix has $\text{det}\Pi \neq 0$, the PS solitons are obtained, as shown in
Figs.~\ref{fig8}($b_{1}$) and ($b_{2}$). The twin peaks of the PS soliton
gradually separate with the increase of $A$.

In this case, the number density is $n(x,t)=4\lambda _{1I}^{2}\text{sech}%
^{2}(2\theta _{R})$ and the total number of atoms is $N_{T}=4\xi _{1I}\,\exp
\left( 2\left( \Omega _{0}/\omega \right) \cos (\omega t)\right) $ when $%
\text{det}\Pi =0$, where $\lambda _{1I}$ is given in Eq.~(\ref{periodic
amplitude}). The total number of atoms $N_{T}$ and soliton's amplitude are
periodic functions of time with period $2\pi /\omega $, as seen in Fig.~\ref%
{fig8}. When $2|\text{det}\Pi |=1$, the number density is $n(x,t)=8\lambda
_{1I}^{2}\text{sech}^{2}(2\theta _{R}-\frac{\ln 2}{2})$ and the total number
of atoms is $N_{T}=8\xi _{1I}\,\exp \left( 2\left( \Omega _{0}/\omega
\right) \cos (\omega t)\right) $, which is twice as large as in the case of $%
\text{det}\Pi =0$. In the same case, the spin density is $\mathbf{f}%
(x,t)=4\lambda _{1I}^{2}\text{sech}^{2}(2\theta _{R})\text{tr}\{\Pi ^{\dag }%
\boldsymbol{\sigma }\Pi \}$, and the total spin $\mathbf{F}_{T}=4\xi
_{1I}\,\exp \left( 2\left( \Omega _{0}/\omega \right) \cos (\omega t)\right)
\text{tr}\{\Pi ^{\dag }\boldsymbol{\sigma }\Pi \}$ with $|\mathbf{F}%
_{T}|=4\xi _{1I}\,\exp \left( 2\left( \Omega _{0}/\omega \right) \cos
(\omega t)\right) $. It is seen that the total number of atoms $N_{T}$ and
total spin $|\mathbf{F}_{T}|$ exhibit the same periodic periodic time
dependence as the soliton's amplitude. When $2|\text{det}\Pi |=1$, the spin
density vanishes everywhere, i.e., $\mathbf{f}(x,t)\equiv (0,0,0)^{T}$, as
it should be in the polar state. When the polarization matrix is constrained
by $2|\text{det}\Pi |<1$, the spin density $n(x,t)$ is not zero but the
total spin still vanished, $\mathbf{F}_{T}\equiv (0,0,0)^{T}$, as shown by
the direct calculation.

\subsubsection{Nonautonomous solitons under the action of sign-reversible
gain-loss distribution}

Considering the evolution of the solitons under the action of the
time-dependent gain-loss term, it is natural to address the case when the
gain and loss are globally balanced, so that the respective coefficient $%
\Gamma (t)$ in Eq. (\ref{vext}) is a localized odd function of time, $\Gamma
(-t)=-\Gamma (t)$. The explore this option, we adopt%
\begin{equation}
\Gamma (t)=W_{0}\mathrm{sech}(\kappa t)\tanh (\kappa t),  \label{pt gain}
\end{equation}%
where $W_{0}$ is a real constant and $\kappa >0$ is the temporal-modulation
parameter. This form of variable coefficient $\Gamma (t)$ resembles the $%
\mathcal{PT}$ symmetry \cite{PT}, but \textquotedblleft rotated" in the $%
\left( x,t\right) $ plane. According to Eq. (\ref{condition}), the
respective time-dependent attractive/expulsive HO-potential strength in the
integrable system is
\begin{eqnarray}
U_{p}(t) &\equiv &\frac{1}{2}\Omega ^{2}(t)=\frac{1}{2}W_{0}\mathrm{sech}%
(\kappa t)\{k\,\mathrm{sech}^{2}(\kappa t)  \notag \\
&&-[\kappa-2W_{0}\mathrm{sech}(\kappa t)]\tanh ^{2}(\kappa t)\}.  \label{pt Up}
\end{eqnarray}%
In this case, we obtain amplitudes of the three components of the
nonautonomous soliton as $A=2|\lambda _{1I}|(|a|,|b|,|c|)^{T}$, where
\begin{equation}\label{pt amplitude}
\begin{aligned}
\lambda _{1I}=&\,\xi _{1I}\,\exp \left( -2\int_{0}^{t}\Gamma (t^{\prime
})\,dt^{\prime }\right)\\
\equiv&\, \xi _{1I}\,\exp \left[ 2\left( W_{0}/\kappa\right)
\text{sech}(\kappa t)\right] .
\end{aligned}
\end{equation}

\begin{figure}[th]
\includegraphics[scale=0.45]{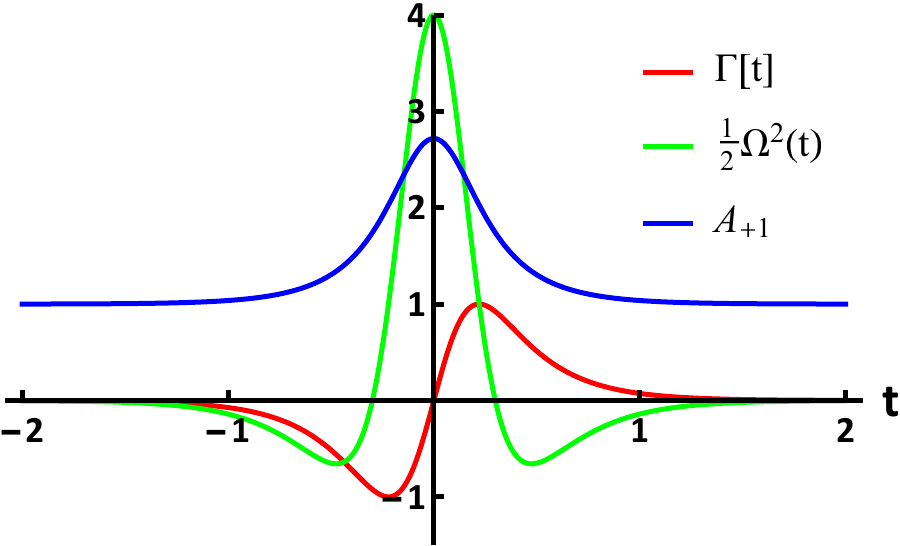}\hspace{1mm} %
\includegraphics[scale=0.45]{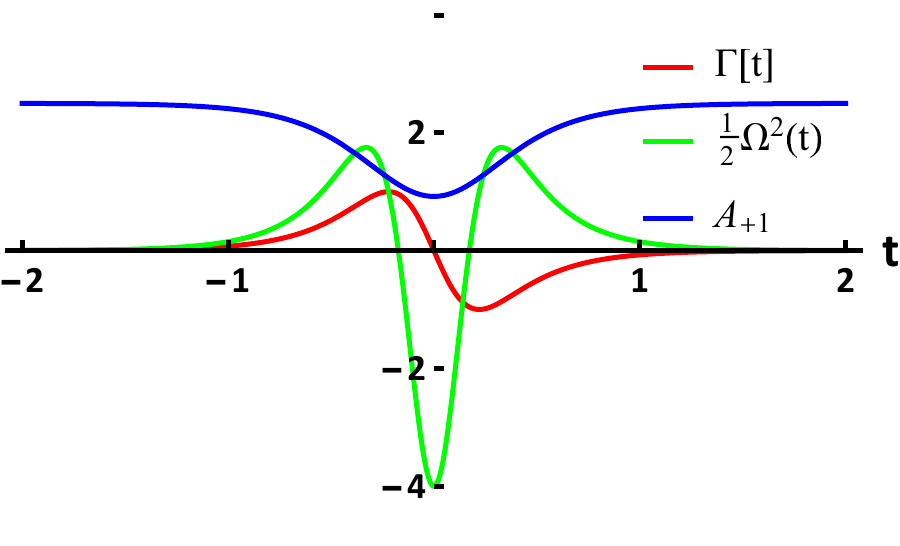} {%
\center{\footnotesize ($a$)
$W_0=2$\hspace{3cm}($b$) $W_0=-2$}}\newline
\caption{The sign-reversible gain-loss coefficient $\Gamma (t)$ given by Eq.
(\protect\ref{pt gain}) (the red lines), strength of the HO potential given
by Eq. (\protect\ref{pt Up}) (the green lines) and amplitude $A_{+1}$ for
component $\protect\phi _{+1}$ (the blue lines), with $W_{0}>0$ and $W_{0}<0$
in $(a)$ and $(b)$, respectively. Other parameters are $\kappa=4$, $a=0.5$, $\protect\xi _{1I}=1$ for $(a)$ and $\protect\xi %
_{1I}=2.5$ for $(b)$.}
\label{fig9}
\end{figure}

According to Eq.~(\ref{pt gain}), the gain-loss coefficient $\Gamma (t)$
varies from $-|W_{0}|/2$ to $+|W_{0}|/2$, attaining these values at $t=\mp
\mathrm{\ln }\left( \sqrt{2}+1\right) /\kappa $, depending on the sign of $%
W_{0}$, as shown by red curves in Fig.~\ref{fig9}. In this solution, the
soliton's amplitude first grows to $2|\xi _{1I}\,a|\exp \left( 2W_{0}/\kappa
\right) $ and then returns to the initial value, $2|\xi _{1I}\,a|$, when $%
W_{0}>0$, or it first decreases to $2|\xi _{1I}\,a|\exp \left( 2W_{0}/\kappa
\right) $ and then returns to the same initial value, $2|\xi _{1I}\,a|$,
when $W_{0}<0$, as shown by blue curves in Fig.~\ref{fig9}. Since the total
gain-loss distribution $\int \Gamma (t)\,dt=0$. In either case, the soliton
recovers to its initial value due to the balance condition, $\int_{-\infty
}^{+\infty }\Gamma (t)dt=0$.

\begin{figure}[th]
\includegraphics[scale=0.45]{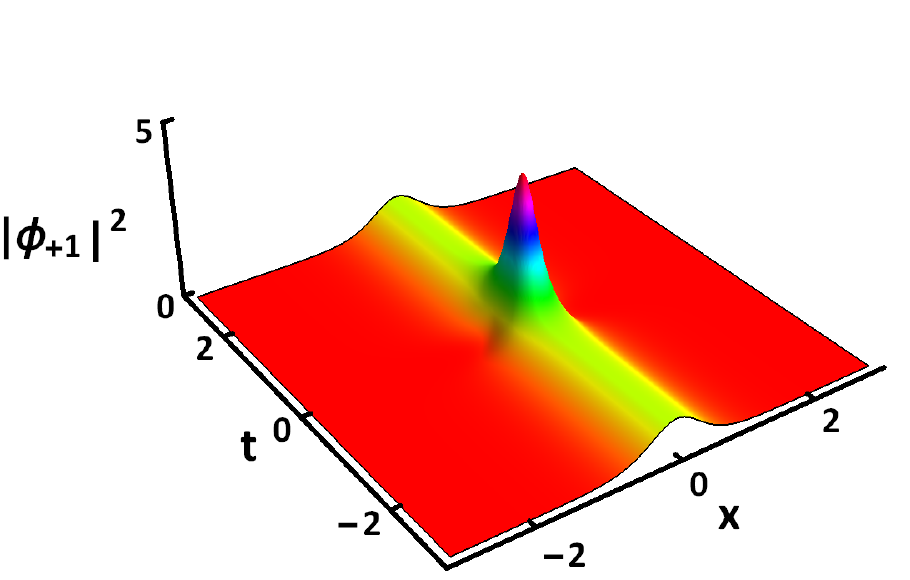}\hspace{1mm} %
\includegraphics[scale=0.45]{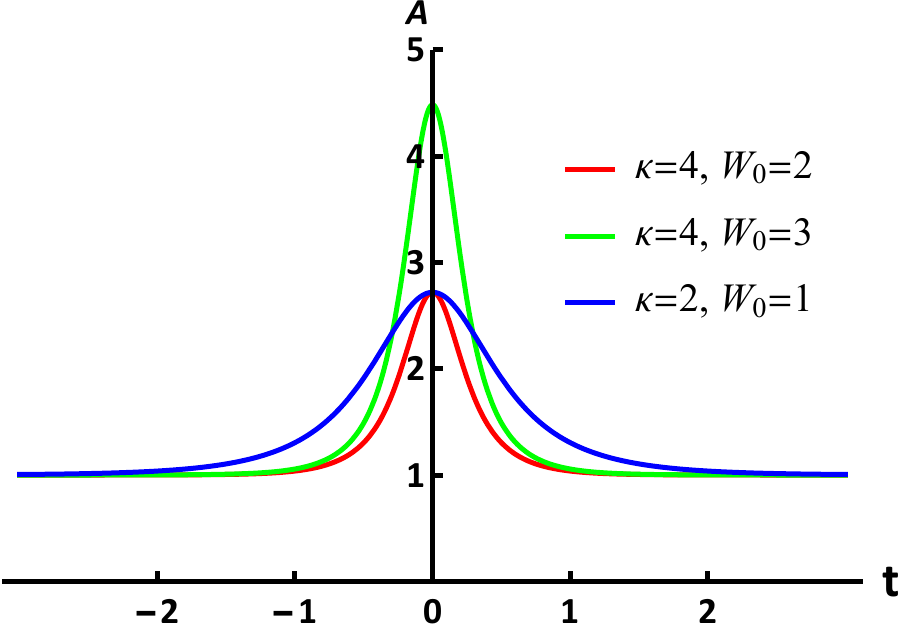} {%
\center{\footnotesize ($a_1$) $A=0
$\hspace{3cm}($a_2$)}}\newline
\includegraphics[scale=0.45]{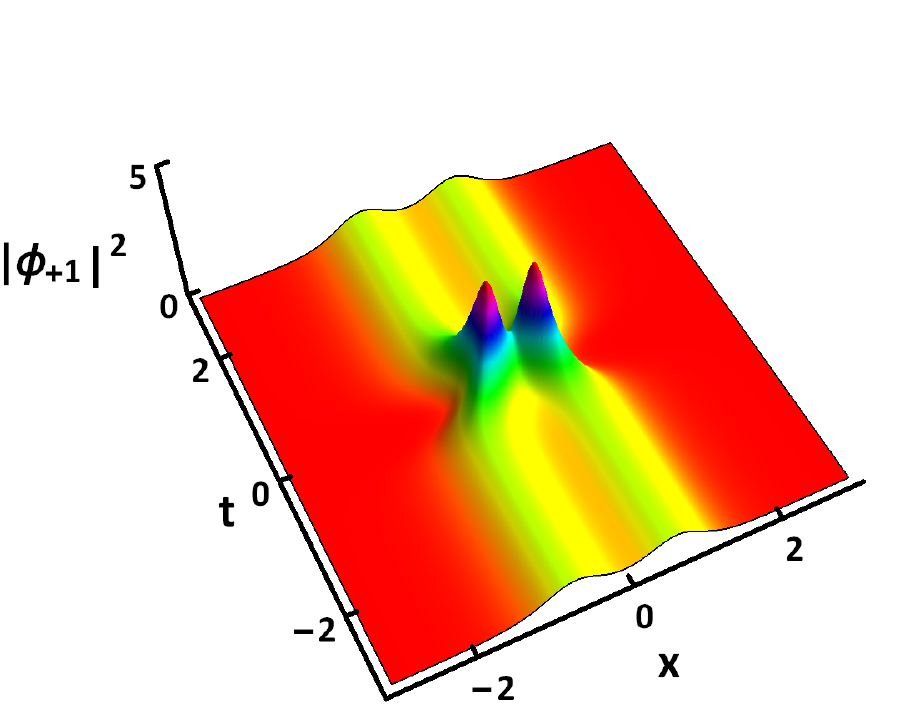}\hspace{1mm} %
\includegraphics[scale=0.45]{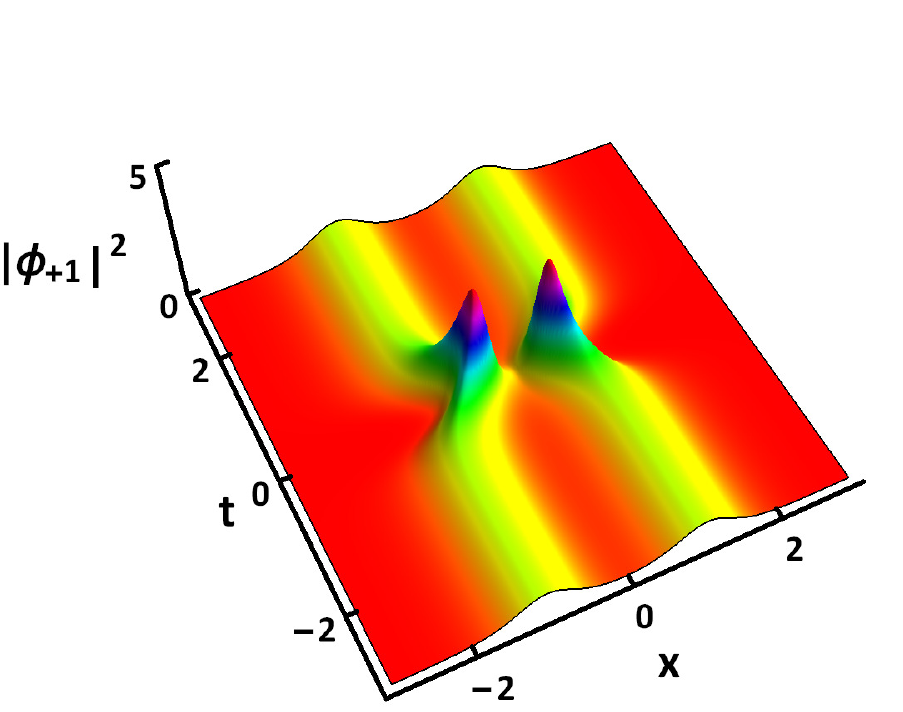} {%
\center{\footnotesize ($b_1$)
$A^2=50$\hspace{2cm}($b_2$) $A^2=1000$}}\newline
\caption{Density profiles of the exact solutions corresponding to Eqs. (%
\protect\ref{pt gain}) and (\protect\ref{pt Up}). ($a_{1}$) The
nonautonomous FS soliton with $\protect\theta _{0}=0$. ($b_{1}$) and ($b_{2}$%
): Nonautonomous double-peak solitons with $\protect\theta _{0}=0.6$ and $%
\protect\theta _{0}=1$, respectively. ($a_{2}$) Effects of the modulation
parameter $\protect\kappa $ and amplitude $W_{0}$ on the steepness and
amplitude of the solitons. Other parameters are $\protect\kappa =4$, $%
W_{0}=1.5$ [except for panel ($a_{2}$)], and $\protect\xi _{1}=\text{i}$.}
\label{fig10}
\end{figure}

The evolution of the exact soliton solution produced by the present setting
is displayed by Fig.~\ref{fig10}, for $W_{0}>0$. It is seen from this figure
and Eq. (\ref{pt amplitude}) that the temporal-modulation parameter $\kappa $
affects the steepness of the arising modulated state and its scaled
amplitude, $W_{0}/\kappa $. These states are similar to rogue waves which
have been widely studied in nonlinear optics, BEC and fluid mechanics~\cite%
{Dudley2019}. In particular, the PS soliton is obtained if the polarization
matrix is restricted to $2|\text{det}\Pi |\leq 1$, as shown in Figs.~\ref%
{fig10}($b_{1}$) and~($b_{2}$). Note that the single-peak states shown in
Figs. \ref{fig10}($a_{1}$) and~($a_{2}$) splits into double-peak ones with
the increase of $A$.

\begin{figure}[th]
\includegraphics[scale=0.45]{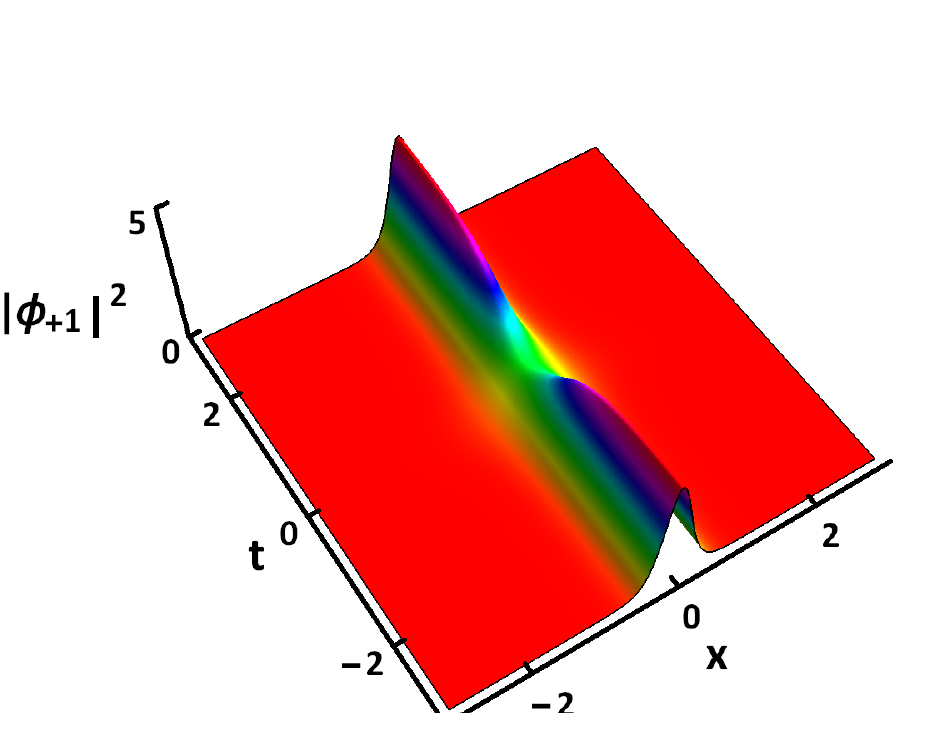}\hspace{1mm} %
\includegraphics[scale=0.45]{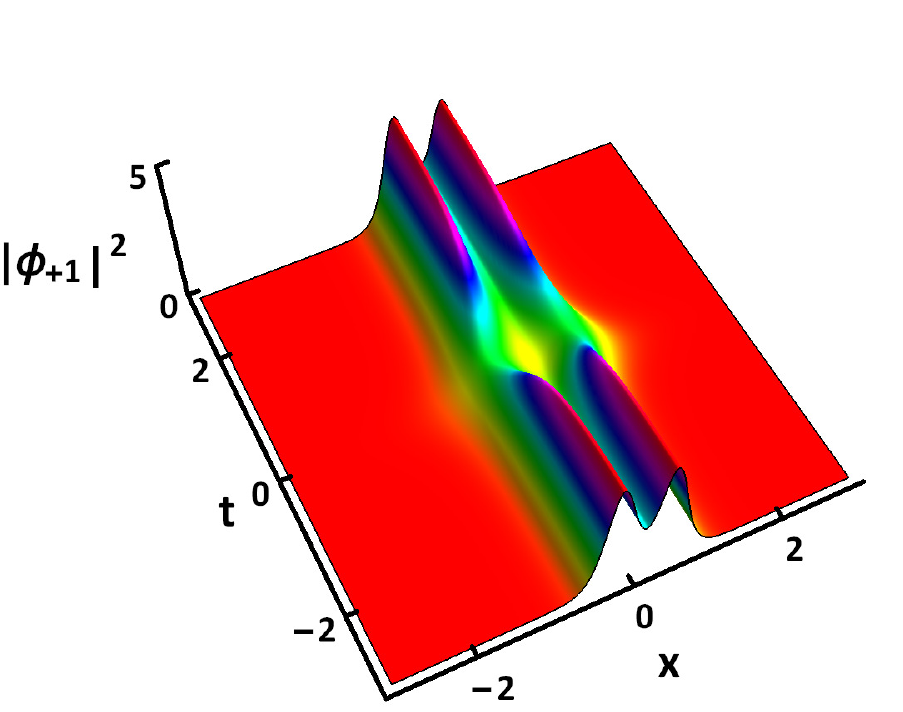} {%
\center{\footnotesize ($a$)
$A=0$\hspace{3cm}($b$) $A^2=50$}}\newline
\caption{Density profiles of solutions of the FS and PS types corresponding
to Eqs. (\protect\ref{pt gain}) and (\protect\ref{pt Up}). The parameters
are $\protect\kappa =2$, $W_{0}=-0.5$, and $\protect\xi _{1}=2\text{i}$.}
\label{fig11}
\end{figure}

In the case of $W_{0}<0$ in Eq. (\ref{pt gain}), the exact solution
produces, instead of the single- and double-peak states in Fig. \ref{fig10},
ones with dips, as shown, for the solutions of both the FS and PS types, as
shown in Fig.~\ref{fig11}.

In this case, we can derive the number density $n(x,t)=4\lambda _{1I}^{2}%
\text{sech}^{2}(2\theta _{R})$ and the respective total number of atoms $%
N_{T}=4\xi _{1I}\exp \left[ 2(W_{0}/\kappa )\mathrm{sech}(\kappa t)\right] $
when $\text{det}\Pi =0$, where $\lambda _{1I}$ is given by Eq.~(\ref{pt
amplitude}). For $2|\text{det}\Pi |=1$, the
atomic-number density is $n(x,t)=8\lambda _{1I}^{2}$sech$^{2}\left[ 2\theta
_{R}-(/2)\ln 2\right] $, and the total number of atoms is $N_{T}=8\xi
_{1I}\,\exp \left[ 2(W_{0}/\kappa )\mathrm{sech}(\kappa t)\right] $, which
is twice that in the case of $\text{det}\Pi =0$. The spin density of this
solution is $\mathbf{f}(x,t)=4\lambda _{1I}^{2}\text{sech}^{2}(2\theta _{R})%
\text{tr}\{\Pi ^{\dag }\boldsymbol{\sigma }\Pi \}$, and the total spin is $%
\mathbf{F}_{T}=4\xi _{1I}\,\exp \left[ 2(W_{0}/\kappa )\mathrm{sech}(\kappa
t)\right] \text{tr}\{\Pi ^{\dag }\boldsymbol{\sigma }\Pi \}$ with $|\mathbf{F%
}_{T}|=4\xi _{1I}\,\exp \left[ 2(W_{0}/\kappa )\mathrm{sech}(\kappa t)\right]
$. When $2|\text{det}\Pi |=1$, the spin density
vanishes like in the cases considered in the above section, therefore the
present solution is also referred to as the PS. When the polarization matrix
is restricted to $2|\text{det}\Pi |<1$, the spin density $n(x,t)$ does not
vanish, but the direct calculation demonstrates that the total spin of the
solution is $\mathbf{F}_{T}(x,t)=0$.

\section{Interaction between nonautonomous matter-wave solitons with
spatiotemporal modulation}

To analyze interactions and collisions between two nonautonomous matter-wave
solitons, one can derive, from the zero seed solution $Q[0]=O$, two matrix
eigenfunctions, $\Psi _{j}=(\mathcal{H}_{j}^{[0]},\mathcal{Y}_{j}^{[0]})^{T}$
($j=1,2$) with
\begin{equation}
\mathcal{H}_{j}^{[0]}=e^{-\theta _{j}}I,~~\mathcal{Y}_{j}^{[0]}=e^{\theta
_{j}}\Pi _{j}^{\ast },  \label{eigenfunction11}
\end{equation}%
where $\theta _{j}=\text{i}\left( \lambda _{j}(t)x+2\int \lambda
_{j}^{2}(t)\,dt\right) +\theta _{0j}$,
\begin{equation}
\Pi _{j}=%
\begin{pmatrix}
a_{j} & b_{j} \\
b_{j} & c_{j}%
\end{pmatrix}%
,  \label{polarization matrix2}
\end{equation}%
\begin{equation}
\begin{aligned} \lambda _{j}(t)=& \left[ \xi _{j}-\frac{1}{2}\int \exp
\left( 2\int \Gamma (t)\,dt\right) \,\gamma (t)dt\right]\times\\
&\exp\left( -2\int \Gamma (t)\,dt\right) \end{aligned}  \label{lambda}
\end{equation}%
(cf. Eq. (\ref{lambda1})), $\xi _{j}$ are complex constants, and $\theta
_{0j}$ are real constants which can be used, as in the case of the single
soliton, to adjust initial positions of the solitons. We also normalize the
matrix $\Pi _{j}$ so that
\begin{equation}
\text{tr}\{\Pi _{j}^{\dag }\cdot \Pi
_{j}\}=|a_{j}|^{2}+2|b_{j}|^{2}+|c_{j}|^{2}=1.  \label{unity2}
\end{equation}

Two-soliton solutions were obtained utilizing the $N$-th-order DT~(\ref{DTN}%
) with $N=2$. The results for collisions displayed below are based on these
solutions. We do not write here the full analytical form of the solutions,
as they are quite ponderous.

To investigate the interactions between two solitons, we first address their
velocities. To this end, we consider the time-independent external potential
$v_{\text{ext}}=\Omega ^{2}x^{2}+\text{i}\Omega $ (see Eq. (\ref{vext})),
with the constant gain-loss coefficient $\Gamma (t)\equiv \Omega $. In this
case, the trajectory of the soliton is determined by equation
\begin{equation}
\xi _{jI}\,e^{-4\Omega t}(\xi _{jR}-\Omega x\,e^{2\Omega t})+\Omega \theta
_{0j}=0.  \label{characteristic line2}
\end{equation}%
Then the velocity of the soliton is derived as $v_{j}=-2\xi
_{jR}\,e^{-2\Omega t}+2\Omega \theta _{0j}\,e^{2\Omega t}/\xi _{jI}$.
Through different choices of polarization matrix $\Pi _{j}$, different types
of solitons can be obtained, such as FS soliton or PS soliton. Two solitons
with different velocities may demonstrate elastic or inelastic interaction,
or form a bound state. Various interaction outcomes can be produced by
altering the polarization matrix $\Pi _{j}$ and velocities of the solitons.

\begin{figure}[th]
\includegraphics[scale=0.45]{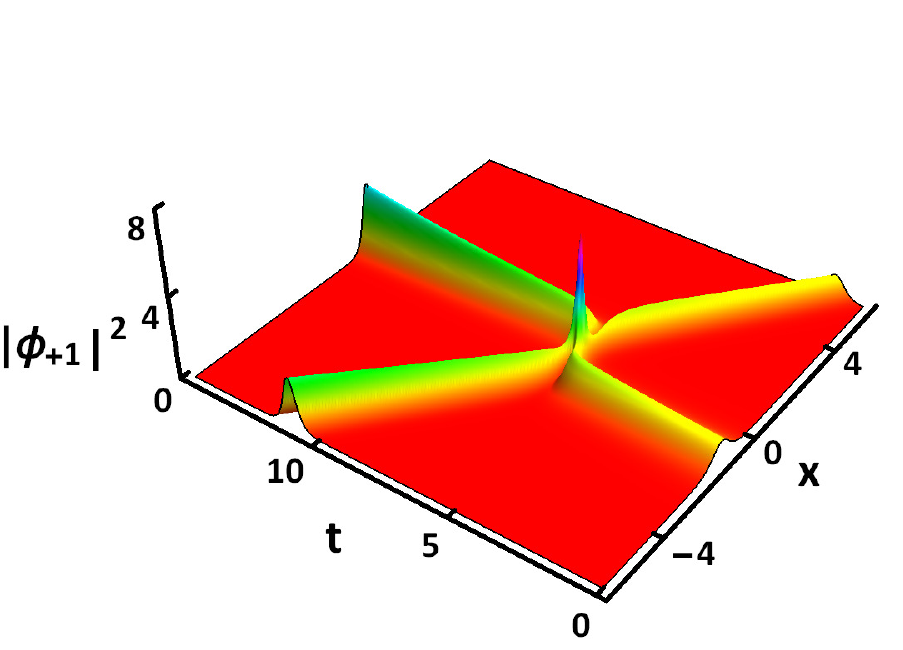}\hspace{1mm} %
\includegraphics[scale=0.45]{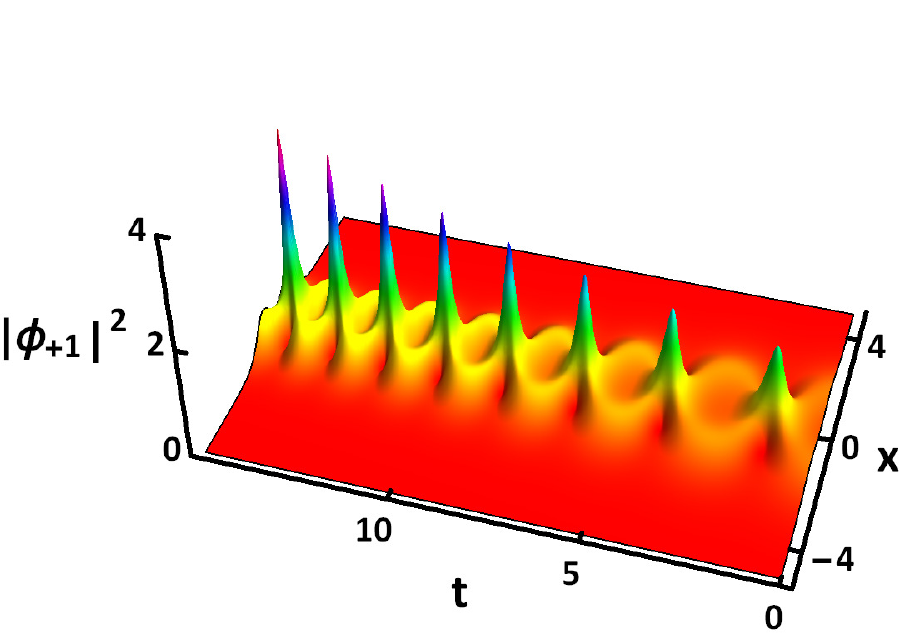} {%
\center{\footnotesize
($a$)\hspace{4cm}($b$)}}\newline
\caption{($a$) The elastic collision between two FS solitons with $\protect%
\xi _{1}=0.2+\text{i}$, $\protect\xi _{2}=-\text{i}$, $\Omega =-0.02$, $%
\protect\theta _{01}=16$ and $\protect\theta _{02}=0$. ($b$) The bound state
of two nonautonomous FS solitons with $\protect\xi _{1}=0.8\text{i}$, $%
\protect\xi _{2}=0.4\text{i}$, $\Omega =-0.02$ and $\protect\theta _{01}=%
\protect\theta _{02}=0$.}
\label{fig12}
\end{figure}

To begin with, we consider the interaction between two FS solitons with $%
\text{det}\Pi _{j}=0$. When they have different velocities, i.e., $v_{1}\neq
v_{2}$, shape-preserving interaction between them is observed, as shown in
Fig.~\ref{fig12}($a$). The only effect of the collision are phase shifts of
the two solitons, which is a typical property of integrable systems.

Then, we address the interaction between two FS solitons possessing the same
velocity $v_{1}=v_{2}$, to generate their bound states. In particular, we
set $v_{1}=v_{2}=0$ by taking $\xi _{jR}=\theta _{0j}=0$, to form the
quiescent bound states of soliton, as shown in Fig.~\ref{fig12}($b$). It is
seen seen that the peak amplitude of the solution exponentially grows under
the action of the gain, $\Gamma (t)\equiv \Omega <0$. At the same time,
period of intrinsic oscillations of the bound state decreases.

\begin{figure}[th]
\includegraphics[scale=0.45]{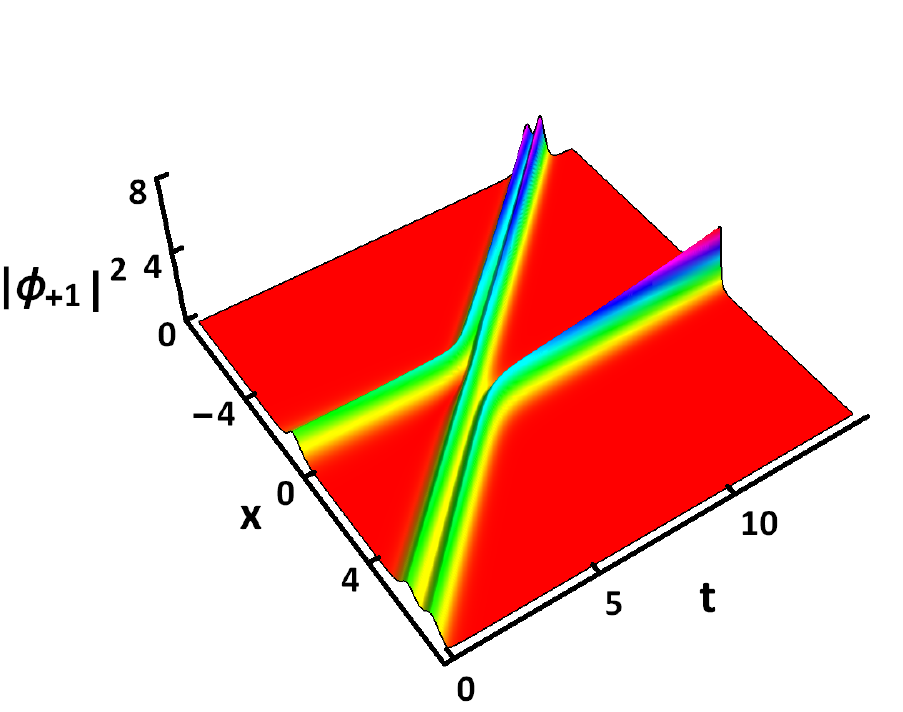}\hspace{1mm} %
\includegraphics[scale=0.45]{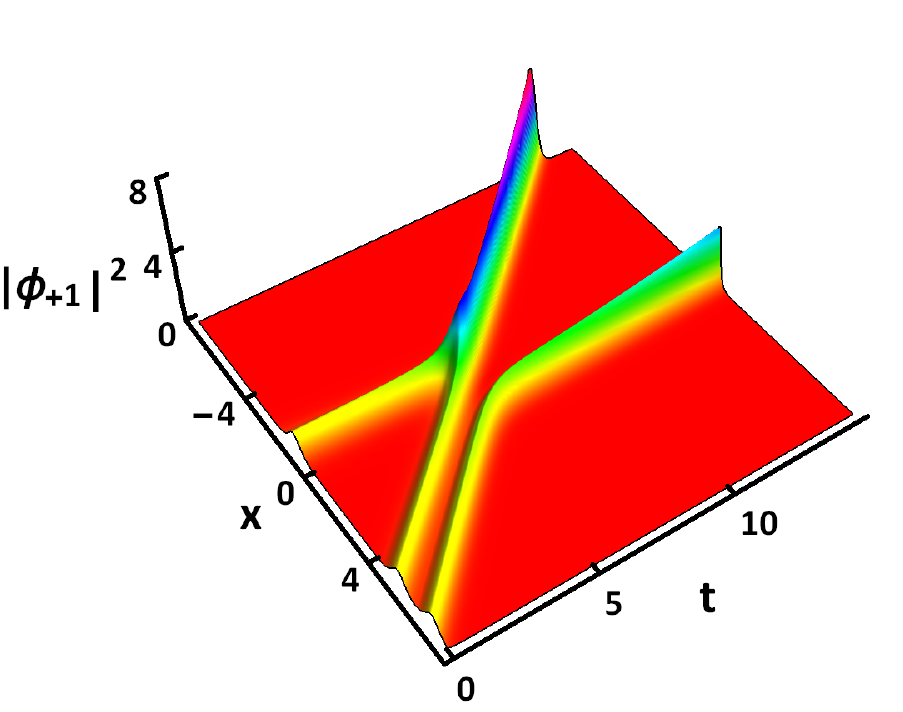} {%
\center{\footnotesize ($a$)
$A_1^2=50$\hspace{2.5cm}($b$) $A_1^2=500$}}\newline
\caption{($a$) The elastic collision between FS and PS solitons. ($b$) The
inelastic collision between FS and PS solitons. The parameters are $\protect%
\xi _{1}=0.2+\text{i}$, $\protect\xi _{2}=\text{i}$, $\Omega =-0.02$, $%
\protect\theta _{01}=16$, and $\protect\theta _{02}=0$.}
\label{fig13}
\end{figure}

Next, we consider the interaction between FS and PS solitons with $\text{det}%
\Pi _{1}\neq 0$ and $\text{det}\Pi _{2}=0$, respectively, i.e., with
different polarization matrices. For instance, with $A_{1}^{2}=4|\text{det}%
\Pi _{1}|^{2}=50$ the interaction is fully elastic, as seen in Fig.~\ref%
{fig13}($a$). As it should be, under the action of the gain the amplitudes
of the solitons grow exponentially. Interestingly, when the determinant of
the polarization matrix $|\text{det}\Pi _{1}|$ decreases, that is, $A_{1}$
increases, inelastic interaction between the PS soliton and FS soliton
occurs. For $A_{1}^{2}=500$, an example is shown in Fig.~\ref{fig13}($b$).
It is seen that the PS soliton changes into a single-hump soliton after the
interaction with a FS soliton. In either case, the FS soliton remains
unchanged after the collision.

\begin{figure}[ht!]
\includegraphics[scale=0.45]{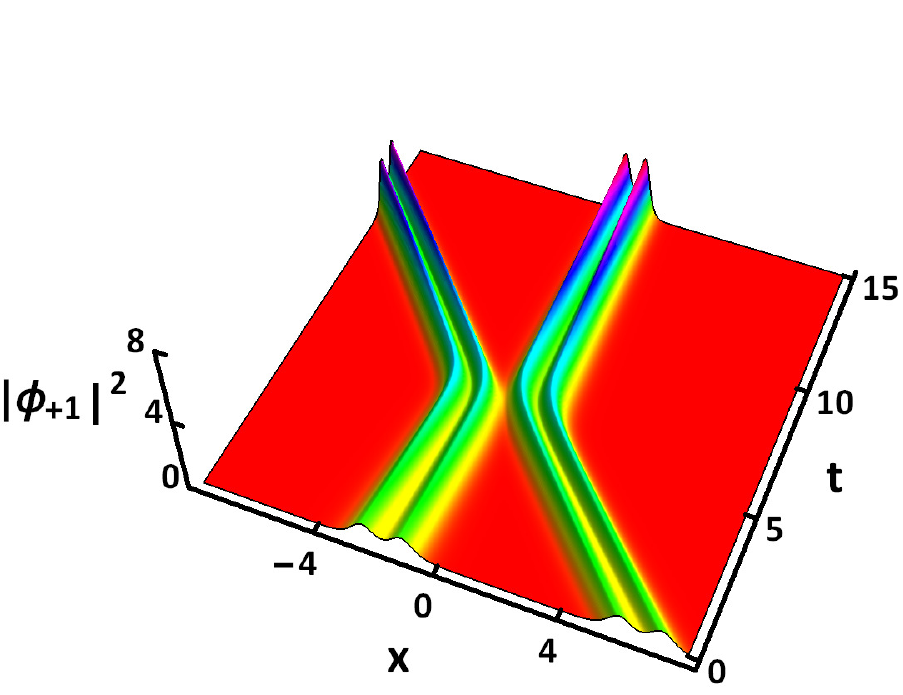}\hspace{1mm} %
\includegraphics[scale=0.45]{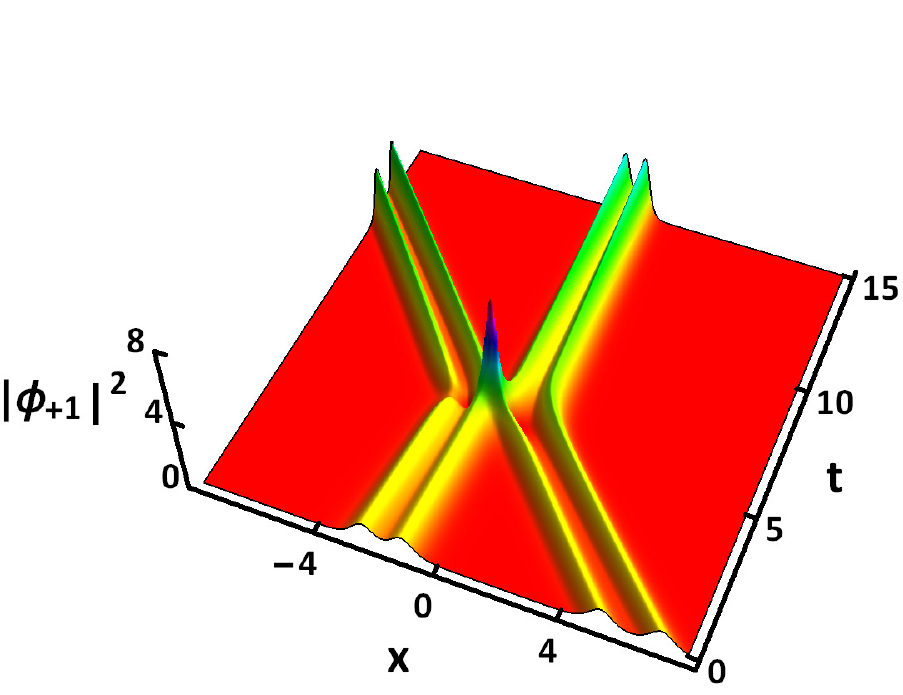} {%
\center{\footnotesize ($a$)
$A_1^2=50$\hspace{2.5cm}($b$) $A_1^2=500$}}\newline
\caption{Shape-preserved interactions between two PS solitons with ($a$) $%
A_2^2=50$ ($b$) $A_2^2=500$. The parameters are chosen as $\protect\xi_1=0.2+%
\text{i}$, $\protect\xi_2=\text{i}$, $\Omega=-0.02$, $\protect\theta_{01}=16$
and $\protect\theta_{02}=0$.}
\label{fig14}
\end{figure}

Finally, we display the elastic collisions between two PS solitons in Fig.~%
\ref{fig14}. It is seen that the two PS solitons keep their double-peak
shapes intact after the interaction. Different from the interaction in Fig.~%
\ref{fig12}($a$), no tall peak is observed in the interaction region in Fig.~%
\ref{fig14}($a$). A general conclusion is that the polarization matrix $\Pi $
can control the strength of the inter-soliton interactions.

\section{Conclusions}

We have investigated the dynamics of nonautonomous matter-wave solitons in
the spinor Bose-Einstein condensate subject to the spatiotemporal
modulation. The model is based on the system of three nonlinearly coupled GP
(Gross-Pitaevskii) equations with the time-dependent potential and gain-loss
coefficient. We have derived the nonisospectral Lax pair with the
time-dependent spectral parameter for this system, provided that the special
integrability condition holds, given by Eq. (\ref{condition}). An infinite
set of conservation laws is derived for the integrable system. Based on the
Lax pair, the DT (Darboux transform) has been constructed and applied to
generate one- and two-soliton solutions. By choosing several different
external potentials and gain-loss coefficients, we have obtained various
solutions for nonautonomous matter-wave solitons of both the ferromagnetic
and polar types (with nonzero and zero total spin, respectively). These
include the compressed, snakelike, and step-wise solitons, as well as
breathers. We have utilized numerical simulation to analyze stability of the
nonautonomous matter-wave solitons against random perturbations and found
that both the ferromagnetic and polar solitons are stable. In particular,
the evolution of the matter-wave solitons, resembling the creation of rogue
waves, has been investigated under the action of the sign-reversible
gain-loss distribution. We have also investigated elastic and inelastic
collisions between nonautonomous matter-wave solitons, including
ferromagnetic-ferromagnetic, ferromagnetic-polar, and polar-polar
collisions. In particular, spin switching has been observed in the inelastic
ferromagnetic-polar collisions. When the two solitons move at the same
velocity, bound states of solitons have been obtained. The outcome of the
interactions can be controlled by the solitons' polarization matrices. Since the integrability condition of system (\ref{system}) have been obtained, the dark non-autonomous solitons of system (\ref{system}) with the repulsive interactions can also be derived via certain analytical method, and the results will be published elsewhere.

\begin{acknowledgments} We express our sincere thanks to all the members of
our discussion group for their valuable comments. This work has been
supported by the National Natural Science Foundation of China under Grant
No. 1197517 and 12261131495, and by the Israel Science Foundation through grant No. 1695/22.
\end{acknowledgments}


\begin{thebibliography}{99}
\bibitem{Anderson1995}
Anderson M. H., Ensher J. R., Matthews M. R., Wieman C. E., Cornell E. A., Observation of Bose-Einstein condensation in a dilute atomic vapor, Science 1995; 269:198.

\bibitem{Hulet}
Bradley C. C., Sackett C. A., Tollett J. J., Hulet R. G., Evidence of Bose-Einstein condensation in an atomic gas with attractive interactions, Phys. Rev. Lett. 1997;79:1170.

\bibitem{Ketterle}
Mewes M. O., Andrews M. R., Druten N. J., Kurn D. M., Durfee D. S., Townsend C. G., Ketterle W., Collective Excitations of a Bose-Einstein Condensate in a Magnetic Trap, Phys. Rev. Lett. 1996;77:988.

\bibitem{Pitaevskii2016}
Pitaevskii L., Stringari S., Bose-Einstein Condensation, Oxford University Press; 2016.

\bibitem{Malomed2022}
Malomed B. A., Multidimensional Solitons, AIP Publishing; 2022.

\bibitem{Kengne2021}
Kengne E., Liu W. M., Malomed B. A., Spatiotemporal engineering of matter-wave solitons in Bose–Einstein condensates, Phys. Rep. 2021;899:1.

\bibitem{ZhangChen2021}
Zhang Z., Chen L., Yao K. X., Chin C., Transition from an atomic to a molecular Bose–Einstein condensate, Nature 2021;592:708.

\bibitem{Musolino2022}
Musolino S., Kurkjian H., Van Regemortel M., Wouters M., Kokkelmans S., Colussi V. E., Bose-Einstein condensation of Efimovian triples in the unitary Bose gas, Phys. Rev. Lett. 2022;128:020401.

\bibitem{Henderson2022}
Henderson G. W., Robb G. R. M., Oppo G. L., Alison M. Yao, Control of light-atom solitons and atomic transport by optical vortex beams propagating through a Bose-Einstein Condensate, Phys. Rev. Lett. 2022;129:073902.

\bibitem{Pethick}
Pethick C. J., Smith  H., Bose-Einstein Condensation in Dilute Gases, Cambridge University Press; 2008.

\bibitem{Kawaguchi2012}
Kawaguchi Y., Ueda M., Spinor bose–einstein condensates, Phys. Rep. 2012;520:253.

\bibitem{Kartashov2019}
Kartashov Y. V., Astrakharchik G. E., Malomed B. A., Torner L., Frontiers in multidimensional self-trapping of nonlinear fields and matter, Nat. Rev. Phys. 2019;1:185.

\bibitem{Malomed2019a}
Malomed B. A., Mihalache D., Nonlinear waves in optical and matter-wave media: a topical survey of recent theoretical and experimental results, Rom. J. Phys. 2019;64:106.

\bibitem{Mihalache2021}
Mihalache D., Localized structures in optical and matter-wave media: a selection of recent studies, Rom. Rep. Phys. 2021;73:403.

\bibitem{LiLi2005}
Li L., Li Z., Malomed B. A., Mihalache D., Liu W. M., Exact soliton solutions and nonlinear modulation instability in spinor Bose-Einstein condensates, Phys. Rev. A 2005;72:033611.

\bibitem{LiZhu2020}
Li H., Zhu X., Malomed B. A., Mihalache D., He Y. J., Shi Z. W., Emulation of spin-orbit coupling for solitons in nonlinear optical media, Phys. Rev. A 2020;101:053816.

\bibitem{Stamper-Kurn2013}
Stamper-Kurn D. M., Ueda M., Spinor Bose gases: Symmetries, magnetism, and quantum dynamics, Rev. Mod. Phys. 2013;85:1191.

\bibitem{Bersano2018}
Bersano T. M., Gokhroo V., Khamehchi M. A., D'Ambroise J., Frantzeskakis D. J., Engels P., and Kevrekidis P. G., Three-Component Soliton States in Spinor $F=1$ Bose-Einstein Condensates, Phys. Rev. Lett. 2018;120:063202.

\bibitem{Evrard2021}
Evrard B., Qu A., Dalibard J., Gerbier F., Observation of fragmentation of a spinor Bose-Einstein condensate, Science 2021;373:1340.

\bibitem{Kim K2021}
Kim K., Hur J., Huh S. J., Choi S., Choi J., Emission of spin-correlated matter-wave jets from spinor Bose-Einstein condensates, Phys. Rev. Lett. 2021;127:043401.

\bibitem{Vengalattore2008}
Vengalattore M., Leslie  S. R., Guzman J., Stamper-Kurn D. M., Spontaneously modulated spin textures in a dipolar spinor Bose-Einstein condensate, Phys. Rev. Lett. 2008;100:170403.

\bibitem{Borgh M O2017}
Borgh M. O., Lovegrove J., Ruostekoski J., Internal structure and stability of vortices in a dipolar spinor Bose-Einstein condensate, Phys. Rev. A 2017;95:053601.

\bibitem{Ollikainen2017}
Ollikainen T., Masuda S., Mottonen M., Nakahara M., Counterdiabatic vortex pump in spinor Bose-Einstein condensates, Phys. Rev. A 2017;95:013615.

\bibitem{Meystre2001}
Meystre P., Atom Optics, Springer-Verlag; 2001.

\bibitem{Peter}
Chen P. Y. P., Malomed B. A., Stable circulation modes in a dual-core matter-wave soliton laser, J. Phys. B: At. Mol. Opt. Phys. 2006;39:2803.

\bibitem{Sekh G A2015}
Sekh G. A., Pepe F. V., Facchi P., Pascazio S., Salerno M., Split and overlapped binary solitons in optical lattices, Phys. Rev. A 2015;92:013639.

\bibitem{Chai2020a}
Chai X., Lao D., Fujimoto K., Hamazaki R., Ueda M., Raman C., Magnetic solitons in a spin-1 Bose-Einstein condensate, Phys. Rev. Lett. 2020;125:030402.

\bibitem{Chai2021a}
Chai X., Lao D., Fujimoto K., Raman C., Magnetic soliton: From two to three components with SO(3) symmetry, Phys. Rev. Res. 2021;3:L012003.

\bibitem{Zhang2009}
Zhang X. F., Hu X. H., Liu X. X., Liu W. M., Vector solitons in two-component Bose-Einstein condensates with tunable interactions and harmonic potential, Phys. Rev, A 2009;79:033630.

\bibitem{Rajendrana2010}
Rajendran S., Muruganandam P., Lakshmanan M., Bright and dark solitons in a quasi-1D Bose–Einstein condensates modelled by 1D Gross–Pitaevskii equation with time-dependent parameters, Physica D 2010;239:366.

\bibitem{Yao Y Q2018}
Yao Y. Q., Han W., Li J., Liu W. M., Localized nonlinear waves and dynamical stability in spinor Bose–Einstein condensates with time–space modulation, J. Phys. B 2018;51:105001.

\bibitem{Atre2006}
Atre R., Panigrahi P. K., Agarwal G. S., Class of solitary wave solutions of the one-dimensional Gross-Pitaevskii equation, Phys. Rev. E 2006;73:056611.

\bibitem{Frantz}
Kevrekidis P. G., Theocharis G., Frantzeskakis D. J. Malomed B. A., Feshbach resonance management for Bose-Einstein condensates, Phys. Rev. Lett. 2003;90:230401.

\bibitem{Yan2013}
Yan M., DeSalvo B. J., Ramachandhran B., Pu H., Killian T. C., Controlling condensate collapse and expansion with an optical Feshbach resonance, Phys. Rev. Lett. 2013;110:123201.

\bibitem{Enomoto2008}
Enomoto K., Kasa K., Kitagawa M., Takahashi Y., Optical Feshbach resonance using the intercombination transition, Phys. Rev. Lett. 2008;101:203201.

\bibitem{Shen2014}
Shen Y. J., Gao Y. T., Zuo D. W., Sun Y. H., Feng Y. J., Xue L., Nonautonomous matter waves in a spin-1 Bose-Einstein condensate, Phys. Rev. E 2014;89:062915.

\bibitem{Malomed2006}
Malomed B. A., Soliton Management in Periodic Systems, Springer; 2006.

\bibitem{Turitsyn2012}
Turitsyn S. K., Bale B. G., Fedoruk M. P., Dispersion-managed solitons in fibre systems and lasers, Phys. Rep. 2012;521:135.

\bibitem{Serkin2007}
Serkin V. N., Hasegawa A., Belyaeva T. L., Nonautonomous solitons in external potentials, Phys. Rev. Lett. 2007;98:074102.

\bibitem{Rajendran2011}
Rajendran S., Lakshmanan M., Muruganandam P., Matter wave switching in Bose–Einstein condensates via intensity redistribution soliton interactions, J. Math. Phys. 2011;52:023515.

\bibitem{YangZhao2011}
Yang Z. Y., Zhao L. C., Zhang T., Feng X. Q., Yue R. H., Dynamics of a nonautonomous soliton in a generalized nonlinear Schr\"{o}dinger equation, Phys. Rev. E 2011;83:066602.

\bibitem{Wang D S2013}
Wang D. S., Shi Y. R., Chow K. W., Yu Z. X., Li X. G., Matter-wave solitons in a spin-1 Bose-Einstein condensate with time-modulated external potential and scattering lengths, The Eur. Phys. J. D 2013;67:1.

\bibitem{Carr2004}
Carr L. D., Brand J., Spontaneous soliton formation and modulational instability in Bose-Einstein condensates, Phys. Rev. Lett. 2004;92:040401.

\bibitem{Davidson}
Rowen E. E., Bar-Gill N., Pugatch R., Davidson N., Energy-dependent damping of excitations over an elongated Bose-Einstein condensate, Phys. Rev. A 2008;77:033602.

\bibitem{Ieda2004}
Ieda J., Miyakawa T. Wadati M., Exact analysis of soliton dynamics in spinor Bose-Einstein condensates, Phys. Rev. Lett. 2004;93:194102.

%

\bibitem{Serkin2010}
Serkin V. N., Hasegawa A., Belyaeva T. L., Nonautonomous matter-wave solitons near the Feshbach resonance, Phys. Rev. A 2010;81:023610.

\bibitem{Janis2005}
Janis J., Banks M., Bigelow N. P., rf-induced Sisyphus cooling in a magnetic trap, Phys. Rev. A 2005;71:013422.

\bibitem{Gericke2008}
Gericke T., W\"{u}rtz P., Reitz D., Langen T., Ott H., High-resolution scanning electron microscopy of an ultracold quantum gas, Nature Phys. 2008; 4(12): 949.

\bibitem{Wurtz2009}
W\"{u}rtz P., Langen T., Gericke T., Koglbauer A., Ott H., Experimental demonstration of single-site addressability in a two-dimensional optical lattice, Phys. Rev. Lett. 2009; 103(8): 080404.

\bibitem{Ablowitz1973}
Ablowitz M. J., Kaup D. J., Newell A. C., Segur H., Nonlinear-evolution equations of physical significance, Phys. Rev. Lett. 1973;31:125.

\bibitem{Darboux}
Rogers C. and Schief W. K.,B\"{a}cklund and Darboux transformations, Cambridge University Press; 2002.

\bibitem{Cramer rules}
Chen Y. L., Representations and Cramer rules for the solution of a restricted matrix equation, Linear and Multilinear Algebra 1993;35:339.

\bibitem{Sandy}
Luo Z., Liu Y., Li Y., Batle J., Malomed B. A., Stability limits for modes held in alternating trapping-expulsive potentials, Phys. Rev. E 2022;106:014201.

\bibitem{PT}
El-Ganainy R., Makris K. G., Khajavikhan M., Musslimani Z. H., Rotter S., Christodoulides D. N., Non-Hermitian physics and PT symmetry, Nature Phys. 2018;14:11.

\bibitem{Dudley2019}
Dudley J. M., Genty G., Mussot A., Chabchoub A., Dias F., Rogue waves and analogies in optics and oceanography, Nature Rev. Phys. 2019;1:675.

\end{thebibliography}

\end{document}